\documentclass[11pt,]{article}
\usepackage{lmodern}
\usepackage{amssymb,amsmath}
\usepackage{ifxetex,ifluatex}
\usepackage{verbatim}
\usepackage{lscape}
\usepackage{fixltx2e} 
\ifnum 0\ifxetex 1\fi\ifluatex 1\fi=0 
  \usepackage[T1]{fontenc}
  \usepackage[utf8]{inputenc}
\else 
  \ifxetex
    \usepackage{mathspec}
  \else
    \usepackage{fontspec}
  \fi
  \defaultfontfeatures{Ligatures=TeX,Scale=MatchLowercase}
\fi
\IfFileExists{upquote.sty}{\usepackage{upquote}}{}
\IfFileExists{microtype.sty}{%
\usepackage{microtype}
\UseMicrotypeSet[protrusion]{basicmath} 
}{}
\usepackage[margin=1in]{geometry}
\usepackage{hyperref}
\hypersetup{unicode=true,
            pdftitle={},
            pdfauthor={},
            pdfborder={0 0 0},
            breaklinks=true}
\usepackage{graphicx,grffile}
\makeatletter
\def\maxwidth{\ifdim\Gin@nat@width>\linewidth\linewidth\else\Gin@nat@width\fi}
\def\maxheight{\ifdim\Gin@nat@height>\textheight\textheight\else\Gin@nat@height\fi}
\makeatother
\setkeys{Gin}{width=\maxwidth,height=\maxheight,keepaspectratio}
\IfFileExists{parskip.sty}{%
\usepackage{parskip}
}{
\setlength{\parindent}{0pt}
\setlength{\parskip}{6pt plus 2pt minus 1pt}
}
\ifx\paragraph\undefined\else
\let\oldparagraph\paragraph
\renewcommand{\paragraph}[1]{\oldparagraph{#1}\mbox{}}
\fi
\ifx\subparagraph\undefined\else
\let\oldsubparagraph\subparagraph
\renewcommand{\subparagraph}[1]{\oldsubparagraph{#1}\mbox{}}
\fi

\let\rmarkdownfootnote\footnote%
\def\footnote{\protect\rmarkdownfootnote}

\usepackage{titling}

\usepackage{float}
\usepackage[svgnames]{xcolor}
\usepackage{booktabs, multirow}
\usepackage{rotating}
\usepackage{booktabs}
\usepackage[font=small,labelfont=bf]{caption}

\usepackage{makecell} 
\usepackage{mathdots}

\usepackage[round]{natbib}

\def\L{\boldsymbol{L}}

\def\X{\bf X}

\def\W{\boldsymbol{W}}

\def\Z{\bf Z}

\def\X{\boldsymbol{X}}
\def\Z{\boldsymbol{Z}}

\def\F{\boldsymbol{F}}
\def\1{\bf 1}
\def\0{\bf 0}

\newfont{\Sc}{eusm10}

\def\bbeta{\mbox{\boldmath{$\beta$}}}
\def\eepsi{\mbox{\boldmath{$\epsilon$}}}
\def\xxi{\mbox{\boldmath{$\xi$}}}

\def\ppsi{\mbox{\boldmath{$\psi$}}}

\title{Evaluating recent methods to overcome spatial confounding}
\author{Arantxa Urdangarin$^{1,2}$, Tom\'as Goicoa$^{1,2,3}$, Mar\'ia Dolores Ugarte$^{1,2,3*}$\\
\small {\textit{$^1$ Department of Statistics, Computer Science, and Mathematics, Public University of Navarre, Spain.}} \\
\small {\textit{$^2$ INAMAT$^2$ (Institute for Advanced Materials and Mathematics) , Public University of Navarre, Spain.}} \\
\small {\textit{$^3$ Institute of Health Research, IdisNA, Spain.}} \\
\small {$*$\textbf{Corresponding author:} Mar\'ia Dolores Ugarte, Department of Statistics, Computer Science, and} \\
\small {Mathematics, Public University of Navarre, Campus de Arrosadia, 31006 Pamplona, Spain.} \\
\small {E-mail: lola@unavarra.es }}
\date{}

\usepackage{listings}
\begin{document}
\maketitle


\begin{abstract}
The concept of spatial confounding is closely connected to spatial regression, although no general definition has been established. A generally accepted idea of spatial confounding in spatial regression models is the change in fixed effects estimates that may occur when spatially correlated random effects collinear with the covariate are included in the model. Different methods have been proposed to alleviate spatial confounding in spatial linear regression models, but it is not clear if they provide correct fixed effects estimates.
In this article, we consider some of those proposals to alleviate spatial confounding such as restricted regression, the spatial+ model, and transformed Gaussian Markov random fields. The objective is to determine which one provides the best estimates of the fixed effects. Dowry death data in Uttar Pradesh in 2001, stomach cancer incidence data in Slovenia in the period 1995-2001 and lip cancer incidence data in Scotland between the years 1975-1980 are analyzed. Several simulation studies are conducted to evaluate the performance of the methods in different scenarios of spatial confounding. Results reflect that the spatial+ method  seems to provide fixed effects estimates closest to the true value although standard errors could be inflated.
\end{abstract}


\section{Introduction}

Research in spatial and spatio-temporal disease mapping has mainly focused on models for smoothing risks in space and time. The models include spatially and temporally correlated random effects as proxies of spatially and temporally structured unobserved covariates with the goal of discovering spatial patterns and their evolution in time. This information is very valuable in epidemiology and public health to highlight regions with high risk and as a first step to discover potential risk factors that may be related to the response of interest. However, this information is somehow preliminar and currently there is an increasing interest in finding associations between hypothetical risk factors and the phenomenon under study. Including potential risk factors (covariates) in a spatial model allows making inference on the strength of the relationship between the response and the covariate. This is usually known as ecological regression.

Spatial regression models including covariates seem a simpler and intuitive mechanism to account for the variability that can be explained by the covariates and the spatially structured variability that remains unexplained, but they present important challenges that continue unsolved (or at least partially unsolved). The most important one is the so called \lq\lq spatial confounding''. This concept has been commonly used to explain the difference between the fixed effect estimates in spatial models and simpler models like ordinary regression that do not consider spatial correlation \citep[see for example][]{Reich2006,Hodges2010}. However, there is neither a unique general definition of spatial confounding nor a definitive solution. This might be the reason why it has been ignored in practice despite its important implications.

\cite{Clayton1993} comment that when \lq\lq the pattern of variation of the covariate is similar to the disease risk, the location may act as a confounder''. Consequently, we would not be stunned if changes in the fixed effects estimates are observed when a spatial term is included in the regression. This might be one of the first references to spatial confounding. Later, \cite{Zadnik2006} conjecture that the change in fixed effects estimates can be due to collinearity between the fixed effects and the conditional autoregressive (CAR) spatial random effects. This collinearity between the fixed effects and the spatial random effects is probably the definition of spatial confounding in spatial linear models in general, and in disease mapping in particular \citep[see for example][]{Reich2006, Hodges2010, Hughes2013, Hanks2015,Page2017,Adin2021}.

Recently, \cite{Gilbert2021} state that spatial confounding is seldom defined explicitly and they point to four phenomena related to this concept.
Namely, 1) bias in the fixed effect estimates due to unobserved variables with spatial pattern; 2) change in fixed effect estimates due to collinearity between fixed and random effects; 3) bias in the fixed effect estimates due to the use of functions to control for spatial dependence such as Markov Gaussian random fields or splines; 4) the challenge of assessing the effect of a covariate with a smooth spatial distribution.
Although they appear different ideas at first sight, they are closely connected. It is widely accepted that spatial random effects (spatial functions) are introduced in the model to adjust for unobserved covariates and hence improve model fitting. However, they may also compete with the observed covariates and then change the fixed effect estimates as an effect of collinearity. Probably, the main difference between these four ideas may be the area of statistics where they appear. For example the first notion is compatible with the definition of confounding in causal inference and there are some examples in the literature \citep[see for example][]{Papa2019, Schnell2020} where spatial confounding is understood as the presence of unmeasured variables with a spatial structure that influence both, an observed covariate and the outcome of interest. In this paper, and to avoid misleading interpretations, we do not pursue the estimation of causal effects, but a rather modest goal: estimating linear associations between a covariate (potential risk factor) and the response of interest in different (spatial) Poisson regression models. We implicitly assume that spatial random effects are included in the model as an approximation to the overall effect of the unobserved covariates \citep{congdon2013,Marques2022}, and this provokes changes in the fixed effect estimates. Then, we investigate which model provides the estimate of fixed effects closest to the true value.

Research on spatial confounding has been focused on existing spatial models to clarify in which conditions they give valid fixed effects estimates \citep{Pacioreck2010}. Probably, the most extended method for dealing with spatial confounding is restricted spatial regression (RSR) proposed by \cite{Reich2006}. RSR is intended to remove collinearity between the covariate of interest and the spatial random effects by restricting the latter to the orthogonal complement of the space spanned by the fixed effects. Hence, the method preserves the fixed effect estimates obtained in a simple regression model without spatial random effects (henceforth null model). \cite{Reich2006} and \cite{Hodges2010} analyse the association between stomach cancer incidence in Slovenia and a socioeconomic indicator (covariate) and justify the RSR because they observe a big change in the fixed effect estimate when a spatial random effect is included in the model. They also explain how the variance of the fixed effect estimator is inflated in the spatial model with respect to the null model. The variance obtained with the RSR is between the variance of the null model and that of the spatial model. However, RSR has been recently criticised. \cite{Khan2020} show that in linear spatial models with normal responses the variance of the RSR fixed effect estimator is always less than or equal to the variance of the null model and hence RSR leads to too liberal inference. For count data they show through simulations that, in certain scenarios, the null model and RSR perform worse than the spatial model if there is spatial variation not explained by observed covariates. Additionally, \cite{Gilbert2021} affirm that RSR presumes no confounding bias. This can be understood because RSR assigns all the variability in the fixed effects direction to the observed convariate assuming that the rest of variability is orthogonal to the observed covariate. Consequently, RSR does not consider the possibility of unobserved overlapping covariates with the observed one and hence the fixed effect estimate should be equal to the null model. Moreover, for these authors, collinearity between fixed and random effects should not be a problem as we would expect a change in fixed effect estimates if we presume there are unobserved covariates. Consequently, the spatial model would account for confounding bias. However, \cite{Hodges2010} show that even if the unobserved covariates are orthogonal to the observed ones, the random effects still provoke changes in the fixed effects.

In the literature there are other methods to alleviate spatial confounding. For example, \cite{Thaden2018} propose a geoadditive structural equation model (gSEM) based on structural equation techniques to account for spatial dependence in both the response and the covariate.  This method is introduced for Gaussian responses and it is not clear how to extend it to non-normal cases because there are two likelihood functions that are modeled together, one for the covariate and one for the response. Additionally, it requires more than one observation per area, precluding its use in disease mapping. Recently, \cite{Dupont2022} propose a method called spatial+ which is a modification of the spatial model. Spatial+ removes spatial dependence from the covariates by fitting spatial spline models to them. The residuals of these fits are then used as explanatory covariates in the spatial regression model for the outcome. The method seems a promising and simple technique to obtain correct fixed effects estimates. A different approach, based on transformed Gaussian Markov random fields and Gaussian copulas, has been proposed by \cite{Prates2015}. The advantage of the method is that the spatial dependence does not interfere with the fixed effects avoiding spatial confounding. All these methods are not free from inconveniences and the main difficulty is to show when and in what circumstances they alleviate confounding effectively.

The main goal of this work is to assess how well recent methods designed to alleviate spatial confounding estimate the fixed effects when there are additional spatially structured variability unexplained by the observed covariates. In particular, we focus on areal count data. For this aim, we simulate several scenarios using different data generating mechanisms that include one observed covariate and additional variability, and fit the different models to compare the fixed effect estimates. We also use the different approaches to revisit real data. Model fitting and inference are carried out from a full Bayes approach using two main techniques: integrated nested Laplace approximations (INLA) and Markov chain Monte Carlo (MCMC) methods.

The rest of the paper is organized as follows. Section 2 briefly introduces the methods used in this work to alleviate spatial confounding. Section 3 illustrates the methods analysing dowry deaths in Uttar Pradesh registered in 2001 \citep{Vicente2020}, the Slovenian stomach cancer data in the period 1995-2001 \citep{Zadnik2006} and the well known  Scottish lip cancer data during the years 1975-1980 \citep[see for example][]{Breslow1993}. Section 4 is devoted to a vast simulation study. Finally, the paper closes with a discussion.

\section{Methods to alleviate spatial confounding}\label{sec_methods}

Throughout this section we assume a large domain (e.g. a country) divided into $n$ small areas (i.e. provinces or districts) labelled as $i=1, 2, ..., n$. Denote by $Y_{i}$ the number of deaths (or incident cases) in the $i$th small area. Then, conditional on the relative risk $r_{i}$, $Y_{i}$ is assumed to be Poisson distributed with mean $\mu_{i}=e_{i}r_{i}$, where $e_{i}$ represents the number of expected cases for area $i$. That is
\begin{equation*}
	Y_{i}\arrowvert r_{i} \sim Poisson(\mu_{i}=e_{i}r_{i}),\;\;\; \text{and} \;\;\; \log \mu_{i}=\log e_{i} + \log r_{i}.
\end{equation*}

In the following, we review some models for $\log r_{i}$ that have been proposed in the literature to deal with confounding.

\subsection{Spatial model}\label{sec_spatial_model}

Spatial regression models include spatial effects to account for the similarity of nearby observations and hence induce spatial smoothness. In disease mapping, Gaussian Markov random fields (GMRF) are used to model spatial random effects \citep[see for example][]{Rue2005}. In particular conditional autoregressive spatial random effects (CAR) have been broadly adopted to capture the spatial dependence that remains unexplained in the model after accounting for covariates. Here, the vector containing the log risks, $\log \boldsymbol{r}$, is modeled as
\begin{equation} \label{spatial_model}
	\log \boldsymbol{r}=\mathbf{1} _{n}\alpha + \bf{X}\bbeta + \xxi
\end{equation}
where  $\boldsymbol{r}=(r_1, r_2, \dots, r_{n})^{'}$ is the vector of relative risks, $\mathbf{1} _{n}$ is a column vector of ones of length $n$, $\alpha$ can be interpreted as an overall risk, ${\X}=({\X}_1,\ldots, {\X}_p)$ is an $n\times p$ matrix whose columns ${\X}_j$, $j=1,\ldots,p$ are the observed covariates, and $\bbeta=(\beta_1, \beta_2, ..., \beta_{p})^{'}$ is the vector of regression coefficients corresponding to the $p$ observed covariates. Finally, $\xxi = (\xi_1, \xi_2, ..., \xi_{n}){'}$ is the vector of spatial random effects which is assumed to follow an intrinsic conditional autoregressive (ICAR) prior \citep{Besag1974}, that is, an improper distribution with  Gaussian  kernel $p(\xxi) \propto \exp(-\frac{1}{2\sigma_{\xi}^2}\xxi^{'}\boldsymbol{Q}_{\xi} \xxi)$. Here, $\boldsymbol{Q}_{\xi}$ is the neighbourhood matrix defined as $\boldsymbol{Q}_{\xi(ij)}=-1$ if areas $i$ and $j$ are neighbours and 0 otherwise, and $\boldsymbol{Q}_{\xi(ii)}$ is equal to the number of neighbours of the $i$th region. Alternatively, spatial effects can be modelled using a smooth function of the coordinates longitude and latitude, that is
\begin{equation} \label{spatial_splines_model}
	\log \boldsymbol{r}=\mathbf{1} _{n}\alpha + \bf{X}\bbeta + \boldsymbol{f}(\boldsymbol{s}_1,\boldsymbol{s}_2),
\end{equation}
where $(\boldsymbol{s}_1,\boldsymbol{s}_2)$ are the coordinates (longitude and latitude) of the centroid of the small areas, and
$\boldsymbol{f}(\boldsymbol{s}_1,\boldsymbol{s}_2)=(f(s_{11}, s_{12}), f(s_{21}, s_{22}), \dots, f(s_{n1}, s_{n2}))^{'}$ is a smooth function to be estimated using, for example, P-splines with a B-spline basis \citep[see for example][]{Ugarte2010,Ugarte2017,Goicoa2019}.

Ignoring the spatial dependence $\xxi$ or $\boldsymbol{f}$ in (\ref{spatial_model}) and (\ref{spatial_splines_model}) we obtain the null model, that is, the model without spatial effects. In our case, a simple Poisson regression model, i.e.
\begin{equation} \label{null_model}
	\log \boldsymbol{r}=\mathbf{1}_{n}\alpha + \bf{X}\bbeta.
\end{equation}

The null model implicitly assumes that all the variability in the response is explained by the observed covariates and there is no confounding bias due to unobserved covariates. Note that spatial models would lead to a change in the fixed effects estimates in comparison to the null model due to the collinearity between the fixed and the random effects. This alleviates confounding according to \cite{Gilbert2021}. Here we understand collinearity between the fixed and the CAR random effects as a collinearity problem between the covariates with spatial structure and the eigenvector of the CAR precision matrix corresponding to the lowest non-null eigenvalue. For a more explicit reformulation of the spatial model \eqref{spatial_model} highlighting the collinearity issue, see for example \cite{Reich2006} or \cite{Goicoa2018}.

\subsection{Restricted spatial regression model}\label{sec_RSR_model}

Restricted spatial regression (RSR) is probably the most popular method to deal with spatial confounding and was first proposed by \cite{Reich2006} to avoid collinearity between fixed and spatial random effects. These authors studied the association between a socioeconomic indicator and stomach cancer incidence in Slovenia. At first sight, they observed that the standardized incidence ratios (SIR), defined as the number of observed cases in one area divided by the number of expected cases in the same area, and the socioeconomic status exhibited  strong spatial patterns. Moreover, a clear negative association between SIR and the socioeconomic status was detected.
The authors first fitted a Poisson regression model (null model) with the socioeconomic status as a single covariate. Secondly, they fitted a spatial model adding spatial random effects that follow the convolution prior proposed by Besag et al. \citep{Besag1991}. They observed that the estimate of the fixed effect in the null and the spatial model changed dramatically: the posterior mean of the fixed effect changed from $-0.137$ (null) to $-0.022$ (spatial) and the posterior variance changed from $0.0004$ (null) to $0.0016$ (spatial). In the case of the Slovenia data, after including the spatial random effects in the model, the negative association between the socioeconomic indicator and stomach cancer disappeared.

To solve this problem, \cite{Reich2006} proposed restricted spatial regression (RSR), a method that consists of restricting the spatial random effects to the space orthogonal to the fixed effects. For count data, the RSR model is expressed as
\begin{equation} \label{RSR_model}
	\log \boldsymbol{r}=\mathbf{1}_{n}\alpha + \X\bbeta + \hat{\bf{W}}^{-1/2}\L\L^{'}\hat{\bf{W}}^{1/2}\xxi
\end{equation}
where the columns of $\L$ are the eigenvectors having non-null eigenvalues of the projection matrix $\boldsymbol{I}_{n}-\hat{\bf{W}}^{1/2}\X_{*}(\X_{*}^{'}\hat{\bf{W}}\X_{*})^{-1}\X_{*}^{'}\hat{\bf{W}}^{1/2}$, which projects onto the orthogonal space of $\hat{\bf{W}}^{1/2}\X_{*}$ being $\X_{*}=[\mathbf{1}_{n}, \X ]$ and $\W$ a diagonal matrix of weights with $W_{ii}=Var(Y_{i}\,\arrowvert\, \alpha, \bbeta, \xxi)=\mu_i$.
In practice, the matrix $\hat{\bf{W}}$ is obtained by fitting the spatial model \eqref{spatial_model}.
Note that the RSR model \eqref{RSR_model} removes collinearity between the fixed and random effects as the combination of $\hat{\bf{W}}^{1/2}\xxi$ in the span of $\hat{\bf{W}}^{1/2}\X_{*}$ is deleted.

RSR removes collinearity, but all the variability in the direction of the fixed effects is attributed to the observed covariate, consequently it implicitly asumes that there is no unobserved covariate that may produce confounding bias. Then, according to \cite{Gilbert2021}, RSR is not a method to alleviate spatial confounding. Additionally, \cite{Khan2020} and \cite{Zimmerman2021} have demonstrated that in spatial models with normal responses the variances of the fixed effects estimates obtained with RSR are less than or equal to the variances obtained with the null model. Consequently, the credible intervals are narrower leading to small coverage rates and an increase of Type-S error rates.  The Type-S error is the Bayesian analogue to the frequentist Type I error \citep[see for example][]{Hanks2015}. That is, a Type-S error occurs if a 95\% equal-tailed credible interval for the regression parameter does not contain zero when the regression parameter is truly zero.

\subsection{Spatial+ method}

Very recently,  \cite{Dupont2022} have proposed a novel approach to reduce spatial confounding when the covariate of interest is spatially structured. These authors show that the bias in the fixed effect estimate is due to spatial smoothing. The Spatial+ method is a modification of the spatial model and reduces bias by eliminating the spatial dependence of the covariate. The method consists of two steps: first, the spatial dependence of the covariate is removed  through a model that we will  denote as {\bf covariate model}. Second, the spatial model is fitted replacing the covariate by the residuals obtained in the first step. We will call this model {\bf spatial+ final model}.
The authors introduce the method using thin plate splines for the spatial effects in both the covariate model and the spatial+ final model. Here we also deal with the spatial dependence in the covariate model using P-splines or including the eigenvectors of the precision matrix $\boldsymbol{Q}_{\xi}$ corresponding to a specific number of the non-null lowest eigenvalues as covariates in a linear model where the observed covariate is now the response. Note that these eigenvectors (in particular the one corresponding to the lowest non-null eigenvalue) are responsible for the collinearity between the fixed and random effects \citep{Reich2006}.
In more detail, the spatial+ method starts from the spatial model \eqref{spatial_splines_model},

\begin{equation*}
	\log \boldsymbol{r}=\mathbf{1} _{n}\alpha + \bf{X}\bbeta + \boldsymbol{f}
\end{equation*}
where $\boldsymbol{f}$ is a spatial term originally modeled with splines (see Dupont et al, 2022). Given the $j$th covariate $\X_{j}$, $j=1,..., p$, we consider the covariate model


\begin{equation}\label{residual_model}
\tilde{\X}_{j}=\tilde{\ppsi}_{j} + \tilde{\eepsi}_{j}
\end{equation}
where $\tilde{\X}_{j}=\hat{\W}^{1/2}\X_{j}$, $\tilde{\ppsi}=\hat{\W}^{1/2}\ppsi$, $\tilde{\eepsi}_{j}=\hat{\W}^{1/2}\eepsi_{j}$, and $\eepsi_{j} \sim N(\mathbf{0}, \sigma_{\X_{j}}^2 \boldsymbol{I}_n)$. Here, $\sigma_{\X_{j}}$ is the standard deviation of the independent and identically distributed errors in the $j$th covariate model, $\boldsymbol{I}_n$ is an $n\times n$ identity matrix, and $\W$ is the same diagonal matrix of weights from Model \eqref{RSR_model}. Finally, $\ppsi$ are spatial effects that can be modeled in two ways. The first one consists of including the eigenvectors of the precision matrix $\boldsymbol{Q}_{\xi}$ corresponding to the $k$ lowest non-null eigenvalues as covariates, so that model \eqref{residual_model} is a weighted linear regression model. Here we choose $k$ so that it is at least $5\%$ and at most $30\%$ of the total number of eigenvectors. The second option uses P-splines or thin plate splines to model the spatial dependence of the covariate. 

The residuals of each covariate $j$ are $\tilde{\Z}_{j}=\tilde{\X}_{j}-\tilde{\ppsi}_{j}$. Once the weighted residuals are computed, they are transformed to the original scale  $\Z_{j}=\hat{\W}^{-1/2}\tilde{\Z}_{j}$ \citep[see][for details]{Dupont2022}. The residuals $\Z_{j}$ are standardized before including them in the spatial+ model.

Finally, the spatial+ final model is fitted replacing the matrix of covariates $\X$ in (\ref{spatial_splines_model}) by the matrix of residuals $\Z$ as
\begin{equation}\label{spatplus_model}
	\log \boldsymbol{r}=\mathbf{1}_{n}\alpha + \Z\bbeta + \boldsymbol{f}.
\end{equation}
Note that in this paper the spatial term $\boldsymbol{f}$ is modeled using ICAR random effects or using splines.

\subsection{Transformed Gaussian Markov Random Field (TGMRF) model}

Transformed Gaussian Markov Random Fields (TGMRF) were introduced by \cite{Prates2015} and are based on the general Gaussian graphical model proposed by \cite{Dobra2011}. The interpretation of the fixed effects is the same as in the previous methods and the main advantage is that the spatial dependence does not interfere with the fixed effects.

In the previous models (spatial model, RSR, and spatial+ model), the main idea is to connect the covariate and the spatial effects with the relative risks using a given link function $g()$. In our case, $g(\boldsymbol{r})=\log \boldsymbol{r}$. Then, the dependence between the relative risks $r_i$ is induced by the prior distribution of the spatial effects. TGMRF provides an alternative way that specifies any positive continuous distribution for the marginal distributions of the relative risks where the covariate effects are introduced in the parameters of the marginal distribution and the spatial dependence structure is captured thanks to the use of a Gaussian copula. Copulas are functions that join multivariate distribution functions to their one-dimensional marginal distribution functions \citep{Nelsen2006}. Sklar's theorem illustrates the role that copulas play in the relationship between multivariate distribution functions and their univariate margins \citep[see Section 2.3 of][]{Nelsen2006}.

Assuming that areal count data follow a Poisson distribution, the TGMRF model is expressed as,
\begin{equation}\label{TGMRF_model}
\boldsymbol{r} \sim TGMRF(\F, \boldsymbol{\Omega}),
\end{equation}
where $\boldsymbol{r}=(r_1, r_2, \dots, r_{n})^{'}$ is the vector of relative risks, $\F=(F_1, F_2, \dots, F_{n})^{'}$, $F_{i}$ is the marginal distribution of $r_{i}$,  and  $\boldsymbol{\Omega}$ is a correlation matrix that determines the spatial dependence structure in the Gaussian copula.
Details about how the marginal distributions for the relative risks are defined in this work, as well as the way of specifying the spatial correlation matrix $\boldsymbol{\Omega}$ are available in Appendix B. In short, the TGMRF method defines the n-dimensional distribution function of the vector of relative risks $\boldsymbol{r}$, denoted as $H$, in two steps. First, a marginal distribution $F_{i}$ is choosen for each $r_{i}$. Then, the multivariate distribution function of $\boldsymbol{r}$ is defined as
\begin{eqnarray*}
p(r_1 \leq a_1, \,\dots,\, r_{n} \leq a_{n})= H(a_1,\,\dots,\, a_{n} \,\arrowvert\, \boldsymbol{\Omega},\, F_1,\, \dots,\, F_{n})\\
=C(F_1(a_1), \, \dots,\, F_{n}(a_{n}) \,\arrowvert\, \boldsymbol{\Omega})
\end{eqnarray*}
where $C(u_1, \, \ldots, \, u_n \,\arrowvert\, \boldsymbol{\Omega})=\Phi_n(\Phi^{-1}(u_1), \, \ldots,\,\Phi^{-1}(u_n) \,\arrowvert\, \boldsymbol{\Omega}):$$[0,\,1]^{n} \rightarrow [0,\,1]$ is a Gaussian copula, $\Phi_{n}(\cdot)$ is the cumulative distribution function of the multivariate normal distribution $N(\mathbf{0}, \, \boldsymbol{\Omega})$ \citep{Dobra2011}, and $\Phi^{-1}$ is the cumulative distribution function of the standard normal random variable. TGMRFs avoid spatial confounding since the covariates are included in the parameters of the marginal distributions $F_{i}$, and as a second step, the spatial dependence is introduced with the Gaussian copula.

In Poisson models, the most common choice for the marginal distribution of each $r_{i}$ is the Gamma distribution. If the covariates are included in the scale parameter, the marginal distribution $F_{i}$ is of the form
\begin{equation*}
\Gamma(1/ \upsilon, \upsilon \exp(\X_{i, \cdot}\, \bbeta)
\end{equation*}
where $\upsilon >0$ and $\X_{i, \cdot}$ is the i$th$ row of the covariate matrix $\X$. When the covariates are included in the shape parameter, the marginal distribution $F_{i}$ takes the form
\begin{equation*}
\Gamma(\exp(\X_{i, \cdot}\, \bbeta) / \upsilon, \upsilon).
\end{equation*}

The TGMRF model is fitted within a full Bayesian framework using Markov chain Monte Carlo (MCMC) algorithms  to draw samples from the posterior distribution of the parameters of interest. The authors of the method have created an \verb"R" package called \verb"TMGMRF" which implements the TGMRF method using NIMBLE \citep{deValpine2017}, and it is available at https://github.com/DouglasMesquita/TGMRF. The rest of the models are fitted using INLA \citep{Rue2009}. Note that INLA provides posterior distributions of the quantities of interest, but it does not rely on MCMC algorithms, thus reducing computing time.

\section{Real data analyses}

In this section, three real data sets are used for illustration purposes: dowry deaths data in Uttar Pradesh in 2001 \citep[see][]{Vicente2020}, stomach cancer incidence data in Slovenia over the period 1995-2001 \citep{Zadnik2006}, and lip cancer incidence data in Scotland during 1975-1980 \citep{Breslow1993}.

All the methods introduced in Section \ref{sec_methods} are fitted to each dataset to estimate the relationship between the relative risks and the covariate of interest. Namely, the null model, the spatial model, the RSR, the spatial+ model and the TGMRF model. A CAR prior for the spatial random effects has been considered in all models. Additionally, the spatial dependence has been modelled using P-splines in the spatial+ method. Regarding the spatial+ technique, two main different approaches have been considered in the covariate model to remove the spatial dependence. In the first one we fit a linear model where the covariate of interest is the response and the $k$ eigenvectors corresponding to the $k$ lowest non-null eigenvalues of the precision matrix $\boldsymbol{Q}_{\xi}$ are the regressors. In the second one, we model the spatial dependence in the covariate using P-splines or thin plate splines. The number of eigenvectors depends on the dimension of the matrix $\boldsymbol{Q}_{\xi}$, i.e, the size of the map. Here a minimum of 5 eigenvectors have been chosen for all data sets whereas the maximum number ranges between 15 and 40. The spatial dependence in the second step of the spatial+ approach has been modelled using an ICAR prior or P-splines. Finally, we fit TGMRF models with gamma marginal distributions including  the covariates in both, the scale (TGMRF1) and the shape parameter (TGMRF2).  Table \ref{tab_spatial+_real_data} displays the notation of the different proposals for the spatial+ approach depending on how we deal with the spatial dependence in the covariate model and in the spatial+ final model.

\begin{table}[htbp]
\centering
\caption{Different proposals for the spatial+ approach depending on how we deal with the spatial dependence in the covariate model and in the spatial+ final model. The column Covariate model indicates the way we remove the spatial dependence of the covariate and the column Spatial+final indicates how we take account of the spatial dependence in the spatial+ final model.}
\begin{tabular}{ccc}
\hline\\
\textbf{Name} & \textbf{Covariate model} & \textbf{Spatial+ final}  \\
\hline
SpatPlus5 & 5 eigenvectors & ICAR prior \\
SpatPlus10 & 10 eigenvectors & ICAR prior \\
SpatPlus15 & 15 eigenvectors & ICAR prior \\
SpatPlus20 & 20 eigenvectors & ICAR prior \\
SpatPlus30 & 30 eigenvectors & ICAR prior \\
SpatPlus40 & 40 eigenvectors & ICAR prior \\
SpatPlusP1 & P-splines & ICAR prior \\
SpatPlusTP1 & Thin plate splines & ICAR prior \\
SpatPlusP2 & P-splines & P-splines \\
SpatPlusTP2 & Thin plate splines & P-splines \\
\hline   \end{tabular}
\label{tab_spatial+_real_data}
\end{table}%


We fit all the models with \verb"R", version 4.0.4. For the TGMRF models, we ran three MCMC chains for each model with 10000 iterations each discarding the first 2000 as a burn-in period. One out of every 20 iterations was saved leading to a total of 1200 iterations. For these models we use the \verb"TGMRF" package.
The rest of the models were fitted using the \verb"R-INLA" package \citep{LindRue2015} version 21.02.23 (dated 2021-04-08) with the full laplace strategy. As recommended by \cite{Gelman2006}, a vague uniform prior on the standard deviation $\sigma_{\xi}$ was considered in the spatial, the RSR, and the spatial+ model with ICAR spatial random effects. A vague normal prior with mean 0 and precision 0.001 is considered for the regression coefficients.

Regarding the dimension of the spline bases in the spatial+ method, the dimension of the thin plate spline basis is 17 for the Uttar Pradesh and the Scotland data. For the Slovenia data, we use dimension 30 as we have more areas. For the P-splines, a total of 11 internal knots were chosen for the marginal bases (longitude and latitude) leading to bases of dimension 13 for the Uttar Pradesh and Scotland data. For the Slovenia data, 28 internal knots are considered giving rise to bases of dimension 30. Finally, cubic polynomials were chosen for the marginal B-spline bases and a RW2 prior distribution on the unknown coefficients was used.
The \verb"mgcv" package (version 1.8-40) was used to fit the covariate model with thin plate splines in the spatial+ approach \citep{Wood2003}. 
Finally, to compare the models in terms of goodness of fit and complexity, we compute the Watanabe-Akaike Information Criterion, WAIC,  \citep{watanabe2010asymptotic}.

\subsection{Dowry death data in Uttar Pradesh}

Very succinctly, dowry is the amount of money, properties or goods that the bride's family gives to the groom's relatives before or after the marriage. The dowry was first designed to protect women from unfair traditions, but it has evolved to an extortion practice and female exploitation. In brief, the groom or the groom's relatives use physical and psychological violence against the woman as a means to achieve a greater dowry. This violence can be extended over time ending up in the death of the woman. This is known as a dowry death. Although any form of dowry is prohibited in India, it is still a widespread practice in that country. For more precise details about dowry and dowry death, the reader is referred to \cite{Vicente2020}.

\begin{figure}[h]
\centering
\includegraphics[width=0.6\textwidth]{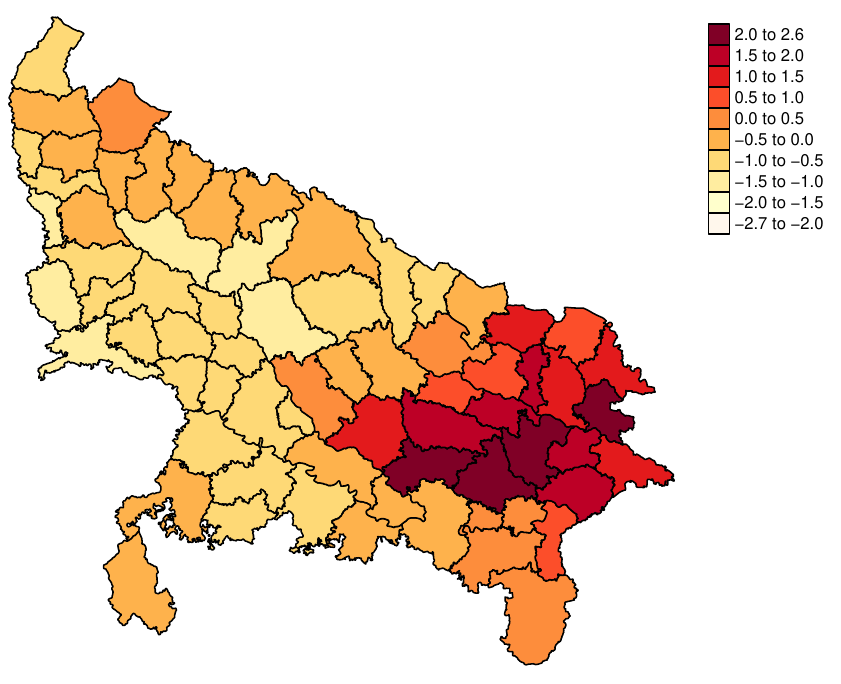}
\caption{Standardized sex ratio covariate in Uttar Pradesh in 2001.}
\label{fig:map_SexRatio}
\end{figure}

In this section, we analyze the number of dowry deaths in 70 districts of Uttar Pradesh in the year 2001. Uttar Pradesh is the Indian state with the highest population and the highest rate of dowry deaths. The goal is to assess if there is a linear association between the covariate sex ratio, defined as the number of females per 1000 males, and the risk of dowry deaths. Figure~\ref{fig:map_SexRatio} shows that the standardized sex ratio has a clear spatial pattern, and hence a collinearity problem with the spatial random effects may exist. Additionally, given the complexity of the dowry death problem, it is very plausible that other unobserved covariates (potential risk factors) may be associated with the dowry deaths and hence confounding bias may appear. Table \ref{tab_results_dowry_death} provides the posterior means of sex ratio, their posterior standard errors, and $95\%$ credible intervals obtained with the different models. The last column of the table shows the WAIC. The differences in the estimates are clear. According to the credible intervals, only two models, the null and the RSR, point towards a significant negative linear association between sex ratio and dowry death relative risk. Spatial and TGMRF models also indicate a negative association, but the 95\% credible intervals contain 0. The rest of models (spatial+ models) provide posterior mean estimates of sex ratio around zero, indicating that the variable is not significant. Regarding standard errors, the TGMRF models lead to higher posterior standard deviations than the spatial models. The spatial+ approach provides posterior standard deviations somewhere in between the null and RSR, and the spatial models.  According to WAIC, all the spatial models but SpatPlusP2 and SpatPlusTP2 lead to similar fits. Clearly, the null model provides the less satisfactory fit.

\begin{table}[htbp]
\centering
\caption{Dowry death analysis in Uttar Pradesh: posterior means of the sex ratio coefficient, posterior standard deviations and $95\%$ credible intervals obtained with different models. The last column shows the WAIC  for each of the models.}
\begin{tabular}{lcccc}
\hline\\
	\textbf{Model} & \textbf{mean} & \textbf{sd} & $\boldsymbol{95\%}$ \textbf{CI}  & \textbf{WAIC} \\
	\hline
	Null & -0.3000 & 0.0238 & (-0.3470, -0.2537)   & 699.4357 \\
	Spatial & -0.0918 & 0.0689 & (-0.2265, 0.0449)  & 463.2098 \\
	RSR & -0.2965 & 0.0226 & (-0.3413, -0.2524)   & 463.1667 \\
	TGMRF1 & -0.1169 & 0.0703 & (-0.2517, 0.0278)    & 461.9326 \\
    TGMRF2 & -0.1004 & 0.0907 & (-0.3020, 0.0716)   & 462.3681 \\
	SpatPlus5 & 0.0012 & 0.0439 & (-0.0855, 0.0875)   & 462.0703 \\
	SpatPlus10 & -0.0092 & 0.0406 & (-0.0892, 0.0705)   & 462.0420 \\
	SpatPlus15 & -0.0160 & 0.0379 & (-0.0908, 0.0582)   & 461.9874 \\
	SpatPlus20 & -0.0227 & 0.0367 & (-0.0952, 0.0493)   & 461.8523 \\
    SpatPlusP1 & 0.0166 & 0.0360 & (-0.0544, 0.0870)   & 462.5284 \\
	SpatPlusTP1 & -0.0043 & 0.0383 & (-0.0801, 0.0706)   & 462.1572 \\
    SpatPlusP2 & 0.0401 & 0.0242 & (-0.0074, 0.0876)   & 495.9132 \\
    SpatPlusTP2 & 0.0239 & 0.0268 & (-0.0281, 0.0771)   & 498.2783 \\
\hline		
\end{tabular}
\label{tab_results_dowry_death}
\end{table}

\subsection{Stomach cancer incidence data in Slovenia}

This data set was first analysed by \cite{Zadnik2006}. The objective is to assess the association between a socioeconomic indicator and the stomach cancer incidence in different regions of Slovenia during the period 1995-2001. \cite{Reich2006} and  \cite{Hodges2010} display the maps of the standardized incidence ratios and the socioeconomic indicator and they observe a negative association.
Table \ref{tab_results_slovenia} shows the posterior mean estimates of the socioeconomic indicator, their posterior standard deviations, and $95\%$ credible intervals as well as the WAIC obtained with the different models. The null model, RSR  and the TGMRF methods estimate a negative regression coefficient for socioeconomic status and the $95\%$ credible interval does not include 0. Otherwise, spatial and spatial+ models estimate regression coefficients very close to 0 and not significant. Similar to the dowry death data, the TGMRF models provides standard errors similar to the spatial model. The null and RSR models lead to the lowest posterior standard deviation, and the spatial+ methods gives posterior standard deviation somewhere in between.  Again, all the spatial models but SpatPlusP2 and SpatPlusTP2 lead to similar fits.

\begin{table}[htbp]
\centering
\caption{Stomach cancer incidence analysis in Slovenia: posterior means of the socioeconomic coefficient, posterior standard deviations and $95\%$ credible intervals obtained with different models. The last column shows the WAIC  for each of the models.}
\begin{tabular}{lcccc}
\hline\\
	\textbf{Model} & \textbf{mean} & \textbf{sd} & $\boldsymbol{95\%}$ \textbf{CI} &   \textbf{WAIC} \\
	\hline
    Null  & -0.1356 & 0.0197 & (-0.1743, -0.0968) &  1146.7657 \\
    Spatial & -0.0351 & 0.0394 & (-0.1116, 0.0431) &  1082.9798  \\
    RSR & -0.1345 & 0.0200  & (-0.1736, -0.0953) &   1082.6928 \\
    TGRMF1 & -0.1028 & 0.0363 & (-0.1695, -0.0335) &     1078.8055 \\
    TGMRF2 & -0.0969 & 0.0377 & (-0.1646, -0.0148) &     1079.9230 \\
    SpatPlus5 & -0.0201 & 0.0282 & (-0.0751, 0.0356) &   1082.3850 \\
    SpatPlus10 & -0.0215 & 0.0283 & (-0.0766, 0.0345) &   1082.4636 \\
    SpatPlus15 & -0.0184 & 0.0288 & (-0.0746, 0.0383) &  1082.1826 \\
    SpatPlus20 & -0.0203 & 0.0285 & (-0.0761, 0.0359) &   1082.1264 \\
    SpatPlus30 & -0.0124 & 0.0290 & (-0.0692, 0.0448) &   1082.2326 \\
    SpatPlus40 & -0.0075 & 0.0273 & (-0.0613, 0.0462) &   1082.0517 \\
SpatPlusP1 & -0.0275 & 0.0289 & (-0.0839, 0.0295) &   1082.3290 \\
SpatPlusTP1 & -0.0150 & 0.0283 & (-0.0704, 0.0408) &  1082.2640 \\
SpatPlusP2 & -0.0418 & 0.0242 & (-0.0890, 0.0058) &   1144.2430 \\
SpatPlusTP2 & -0.0261 & 0.0236 & (-0.0723, 0.0203) &   1145.3580 \\

	\hline	
\end{tabular}
\label{tab_results_slovenia}
\end{table}

\subsection{Lip cancer incidence data in Scotland}

Finally, lip cancer incidence data in Scotland during 1975-1980 is analyzed. A covariate indicating the proportion of the population engaged in agriculture, fishing, or forestry, hereafter named AFF, is included in the models \citep[see for example][]{Breslow1993}. Table \ref{tab_results_scotland} provides the posterior estimates of the regression coefficient of AFF with their posterior standard deviations, $95\%$ credible intervals and WAIC values. The methods estimate a positive regression coefficient for AFF.  However, all the spatial+ models, except the one with 5 eigenvectors as regressors and the ones that model the spatial dependence in the spatial+ final model with splines, provide 95\% credible intervals that include 0, hence discarding an association between AFF and lip cancer incidence relative risks.

\begin{table}[htbp]
\centering
\caption{Lip cancer incidence analysis in Scotland: regression coefficient estimates of AFF with their standard deviations and $95\%$ credible intervals. The last
column shows the WAIC for each of the models.}
\begin{tabular}{lcccc}
\hline\\
	\textbf{Model} & \textbf{mean} & \textbf{sd} & $\boldsymbol{95\%}$ \textbf{CI}  & \textbf{WAIC} \\
	\hline
	Null & 0.5028 & 0.0406 & (0.4228, 0.5822)   & 460.9447 \\
	Spatial & 0.2383 & 0.0881 & (0.0600, 0.4066)  & 294.0840 \\
	RSR & 0.5425 & 0.0447 & (0.4546, 0.6299)   & 294.0726 \\
    TGRMF1 & 0.2529 & 0.0898 & (0.0813, 0.4134)    & 292.8233 \\
    TGMRF2 & 0.1845 & 0.0743 & (0.0438, 0.3277)   & 292.8233 \\
    SpatPlus5 & 0.1650 & 0.0802 & (0.0037, 0.3196)   & 292.7422 \\
    SpatPlus10 & 0.0786 & 0.0751 & (-0.0714, 0.2244)   & 292.4676 \\
    SpatPlus15 & 0.0673 & 0.0787 & (-0.0892, 0.2203)   & 292.5820 \\
    SpatPlusP1 & 0.0944 & 0.0776 & (-0.0613, 0.2440)   & 293.1708 \\
    SpatPlusTP1 & 0.0434 & 0.0780 & (-0.1124, 0.1944)   & 293.0236 \\
    SpatPlusP2 & 0.1425 & 0.0535 & (0.0358, 0.2459)   & 330.4622 \\
    SpatPlusTP2 & 0.1266 & 0.0546 & (0.0180, 0.2325)   & 331.1734 \\
\hline	
\end{tabular}
\label{tab_results_scotland}
\end{table}

In summary, depending on the model used to analyse the data, different estimates of the fixed effects and standard errors are obtained. We note that standard errors seem to be too high in several models. In terms of goodness of fit, the null model presents larger WAIC values than the rest of the methods, so it is not an adequate model for smoothing the risks. Differences among the rest of the models are minor indicating that the procedures lead to a similar smoothing. SpatPlusP2 and SpatPlusTP2 models provide larger values of WAIC than the others. This might probably happens because they oversmooth the risks. Due to the observed discrepancies in the estimates, a simulation study is performed to evaluate which model recovers best the true value of the fixed effects in several scenarios of spatial confounding. Additionally, we also evaluate which model provides appropriate estimates of the standard error.

\section{Simulation study}

In this section, we conduct a complete simulation study to evaluate how the different models estimate the fixed effects in the presence of spatial confounding. For the simulation, we use the geographical setup of Uttar Pradesh consisting of 70 connected districts and the standardized observed covariate sex ratio, denoted as ${\X}_1$. To simulate the log risks, we use ${\X}_1$ and an additional covariate ${\X}_2$ which is generated to have high, intermediate and low correlation with ${\X}_1$. The ${\X}_2$ variable will play the role of an unobserved covariate.

We consider two different scenarios named {\bf Simulation study 1} and {\bf Simulation study 2}.

{\bf Simulation study 1}: The goal of this simulation study is to assess how well the different models estimate the fixed effect ${\X}_1$ when there is spatial confounding. To do this, the data generating model includes both covariates ${\X}_1$ and ${\X}_2$, and additional spatial variability is added in some scenarios. Then we fit the models without the covariate ${\X}_2$. Note that ${\X}_2$ is treated as an unobserved covariate in the fitted models that may produce spatial confounding. In more detail, we first generate the logarithm of relative risks and then we simulate the counts using the Poisson distribution, that is
\begin{eqnarray} \label{eq:simulate_data1}
\log \, \boldsymbol{r}= \X \bbeta + \boldsymbol{S} \label{eq:simulate_data1}\\
\boldsymbol{Y}^{k}\arrowvert\boldsymbol{r} \sim Poisson(\boldsymbol{\mu}=\boldsymbol{er}), \label{eq:simulate_data2}
\end{eqnarray}
where $k=1,\ldots, K$, $\X=({\X}_1, {\X}_2)$, $\boldsymbol{e}$ is the vector of expected cases taken from the real case study (dowry deaths data), and $\bbeta=(\beta_1, \beta_2)^{'}$. Here, $\bbeta=(0.2, 0.3)^{'}$. Note that the generating model includes both covariates ${\X}_1$ and ${\X}_2$ to simulated the log risks. Finally,   $\boldsymbol{S}$ is a term to introduce additional spatial variability. Three different scenarios are considered depending on how we generate the term $\boldsymbol{S}$.
\begin{itemize}
  \item \textbf{Scenario 1}: Here we do not include additional spatial variability. That is $\boldsymbol{S}=\mathbf{0}$.
  \item \textbf{Scenario 2}: The spatial variability is generated using an ICAR model, that is $\boldsymbol{S}={\xxi}$ with $p(\xxi) \propto \exp(-\frac{1}{2\sigma_{\xi}^2}\xxi^{'}\boldsymbol{Q}_{\xi} \xxi)$ and $\sigma_{\xi}^2=0.2$.
  \item \textbf{Scenario 3}: The spatial variability is a smooth surface built using P-splines. That is $\boldsymbol{S}=\boldsymbol{f}(\boldsymbol{s}_{1}, \boldsymbol{s}_{2})=\boldsymbol{B}_{s}\boldsymbol{\theta}$ defined as in \cite{Ugarte2017}, where $\boldsymbol{s}_{1}$ and $\boldsymbol{s}_{2}$ are vectors containing the longitude and latitude of the centroids of the small areas, $\boldsymbol{B}_{s}$ is a two dimensional B-spline basis of dimensions $n\times k_1 k_2$, and $\boldsymbol{\theta}=(\theta_1, \theta_2, \dots, \theta_{k_1 k_2})'$ is the vector of coefficients. Here, the number of elements of the marginal B-splines bases for longitude and latitude is set to $k_1=k_2=13$, leading to 169 elements in the spatial B-spline basis ${\boldsymbol B}_s$.  To generate a smooth surface, the following prior is considered for the coefficients, $\boldsymbol{\theta} \sim N(\boldsymbol{0}, \boldsymbol{P})$, where $\boldsymbol{P}=\lambda_1\boldsymbol{I}_{k_1}\otimes \boldsymbol{D}_1'\boldsymbol{D}_1+\lambda_2 \boldsymbol{D}_2'\boldsymbol{D}_2\otimes\boldsymbol{I}_{k_2}$ is a precision matrix and $\boldsymbol{D}_{1}$ and $\boldsymbol{D}_{2}$ are difference matrices of order 2. Here, different degree of smoothing is considered for longitude and latitude  \citep[see][]{Ugarte2017}. In particular, the hyperparameters that control the amount of smoothing in longitude and latitude are set at $\lambda_{s_1}=1.22$ and $\lambda_{s_2}=8.87$.

\end{itemize}

\begin{figure}[h]
\centering
\includegraphics[width=1\textwidth]{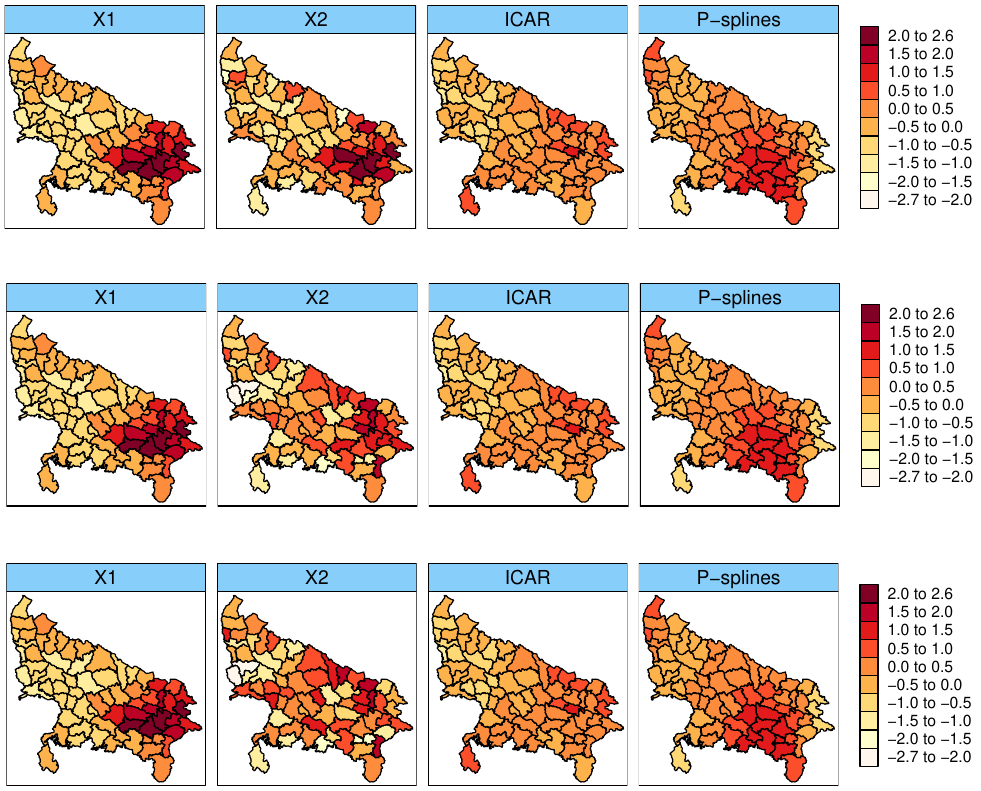}
\caption{From left to right, spatial patterns of the covariate sex ratio (${\X}_1$), the simulated covariate ${\X}_2$, and spatial effects simulated with an ICAR (Scenario 2) or using P-splines (Scenario 3). In the top row $cor({\X}_1,{\X}_2) = 0.8$, in the middle row $cor({\X}_1, {\X}_2)=0.5$, and in the bottom row $cor({\X}_1, {\X}_2)=0.2$.}
\label{fig:maps_covariates}
\end{figure}

For each one of these scenarios, three subscenarios have been generated according to a high, medium or low correlation between the covariates ${\X}_1$ and ${\X}_2$. Namely, subscenario 1 with $cor({\X}_1, {\X}_2)=0.8$, subscenario 2 with $cor({\X}_1, {\X}_2)=0.5$, and subscenario 3 with $cor({\X}_1, {\X}_2)=0.2$. Figure \ref{fig:maps_covariates} displays the spatial patterns of the covariates, the ICAR and the smooth spatial surfaces. The first row shows the spatial patterns of the covariates when the correlation is 0.8. The second row shows the spatial patterns of the covariates when the correlation is 0.5, and the third row corresponds to correlation 0.2. Note that the ICAR and the smooth spatial pattern are simulated only once and they are the same in the three rows. The correlations between sex ratio and the spatial effects simulated with an ICAR or using P-splines are $cor({\X}_1, \xxi)=0.5865$ and $cor({\X}_1, f(\boldsymbol{x}_1, \boldsymbol{x}_2))=0.1998$ respectively. In total we have 9 scenarios, and for each one we generate $K=100$ data sets. Table \ref{tab_scenarios_simu_1} summarizes the details of all the scenarios in Simulation Study 1.

\begin{table}[htbp]
\centering
\caption{Different scenarios considered in Simulation study 1 depending on the data generating model. The second column indicates the data generating model in each of the scenarios. The third column specifies the correlation between $\X_1$ and $\X_2$ in each of the subscenarios.}
\begin{tabular}{lcc}
\hline
&  & Subscenario 1  \\
\vspace{0.15cm}
&  & $cor({\X}_1, {\X}_2)=0.8$ \\
& \textbf{Scenario 1} & Subscenario 2 \\
\vspace{0.15cm}
& $ \text{log} \, \boldsymbol{r}= \X \bbeta$ &  $cor({\X}_1, {\X}_2)=0.5$ \\
& & Subscenario 3 \\
\vspace{0.15cm}
& & $cor({\X}_1, {\X}_2)=0.2$ \\
\cline{2-3}
&  & Subscenario 1  \\
\vspace{0.15cm}
&  & $cor({\X}_1, {\X}_2)=0.8$ \\
\multirow{2}[2]{*}{\textbf{Simulation study 1}\;\;\;\;} & \textbf{Scenario 2} & Subscenario 2 \\
\vspace{0.15cm}
& $\text{log} \, \boldsymbol{r}= \X \bbeta + \xxi $ &  $cor({\X}_1, {\X}_2)=0.5$ \\
& & Subscenario 3 \\
\vspace{0.15cm}
& & $cor({\X}_1, {\X}_2)=0.2$ \\
\cline{2-3}
&  & Subscenario 1  \\
\vspace{0.15cm}
&  & $cor({\X}_1, {\X}_2)=0.8$ \\
& \textbf{Scenario 3} & Subscenario 2 \\
\vspace{0.15cm}
& $\text{log} \, \boldsymbol{r}= \X \bbeta + \boldsymbol{B}_{s}\boldsymbol{\theta} $ &  $cor({\X}_1, {\X}_2)=0.5$ \\
& & Subscenario 3 \\
\vspace{0.15cm}
& & $cor({\X}_1, {\X}_2)=0.2$ \\
\hline
\end{tabular}
\label{tab_scenarios_simu_1}
\end{table}

{\bf Simulation study 2}: The goal of this simulation study is to assess Type-S error rates to complement the information in Simulation study 1.
In this simulation study the log risks are simulated using ${\X}_1$ and additional spatial variability. Then, the models are fitted including ${\X}_2$ to see if any of the models tend to identify this covariate as significant when in fact it is not part of the generating model. The generating process is similar to the one in Simulation study 1, but now $\beta_2=0$ to remove the covariate ${\X}_2$.

All the methods introduced in Section \ref{sec_methods} are fitted to the simulated data. The goal of the study is to assess how well the different methods recover the true value of the fixed effect coefficient and how the posterior standard deviation approximates the true standard error of the estimator. In addition, a method with low Type-S error rates is preferred. Regarding TGMRF approach, both TGMRF1 and TGMRF2 provide pretty similar results, so to conserve space we only report on TGMRF1.

\subsection{Simulation study 1: Results}

The goal of the simulation study is two-fold. On the one hand we evaluate how well the different methods estimate the fixed effects, something crucial to establish the linear relationship between the response and the covariates. On the other hand, we also investigate if the models recover the true risk surface, something relevant to identify potential risk factors.

\begin{table}[htbp]
  \centering
  \caption{Posterior means and standard deviations of $\beta_1$ based on 100 simulated datasets for Simulation study 1, Scenarios 1, 2 and 3 and $cor({\X}_1, {\X}_2)=0.8$, $0.5$ and $0.2$.}
  \resizebox{\textwidth}{!}{
    \begin{tabular}{ccccccccc}
    \hline\\
          &          &   & \multicolumn{2}{c}{\textbf{cor=0.80}}  & \multicolumn{2}{c}{\textbf{cor=0.50}}  & \multicolumn{2}{c}{\textbf{cor=0.20}} \\
    \midrule
    & \textbf{Model} & \textbf{True value} & \textbf{mean}  & \textbf{sd}  & \textbf{mean}  & \textbf{sd}  & \textbf{mean}  & \textbf{sd} \\
    \midrule
    \multirow{12}[2]{*}{\textbf{Scenario 1}} & Null & \multirow{12}[2]{*}{0.2000} & 0.4551 & 0.0160 & 0.3277 & 0.0168 & 0.2353 & 0.0175 \\
    & Spatial & & 0.4603 & 0.0448 & 0.3240 & 0.0593 & 0.2299 & 0.0665 \\
    & RSR & & 0.4569 & 0.0160 & 0.3277 & 0.0171 & 0.2343 & 0.0178 \\
    & TGRMF1 & & 0.4446 &	0.0313 & 0.3383	& 0.0449 & 0.2460 &	0.0505 \\
    & SpatPlus5 & & 0.2838 & 0.0377 & 0.1940 & 0.0407 & 0.1359 & 0.0429 \\
    & SpatPlus10 & & 0.2284 & 0.0438 & 0.1920 & 0.0398 & 0.1443 & 0.0403 \\
    & SpatPlus15 & & 0.1861 & 0.0410 & 0.1436 & 0.0386 & 0.1029 & 0.0383 \\
    & SpatPlus20 & & 0.1486 & 0.0401 & 0.1389 & 0.0361 & 0.1077 & 0.0359 \\
    & SpatPlusP1 & & 0.1177 & 0.0386 & 0.0986 & 0.0361 & 0.0763 & 0.0356 \\
    & SpatPlusTP1 & & 0.1579 & 0.0409 & 0.1350 & 0.0378 & 0.1054 & 0.0376 \\
    & SpatPlusP2 & & 0.1174 & 0.0206 & 0.0944 & 0.0213 & 0.0725 & 0.0218 \\
    & SpatPlusTP2 & & 0.1802 & 0.0261 & 0.1341 & 0.0267 & 0.0995 & 0.0281 \\
    \midrule
    \multirow{12}[2]{*}{\textbf{Scenario 2}} & Null & \multirow{12}[2]{*}{0.2000} & 0.6612 & 0.0144 & 0.5398 & 0.0149 & 0.4564 & 0.0154 \\
    & Spatial & & 0.5683 & 0.0834 & 0.4389 & 0.0900 & 0.3440 & 0.0949 \\
    & RSR & & 0.6528 & 0.0152 & 0.5273 & 0.0160 & 0.4444 & 0.0165 \\
    & TGRMF1 & & 0.6245 &	0.0840 & 0.4875 & 0.0951 & 0.3965 & 0.1001 \\
    & SpatPlus5 & & 0.3468 & 0.0627 & 0.2582 & 0.0627 & 0.1954 & 0.0630 \\
    & SpatPlus10 & & 0.2579 & 0.0699 & 0.2429 & 0.0642 & 0.2001 & 0.0625 \\
    & SpatPlus15 & & 0.2257 & 0.0603 & 0.1958 & 0.0561 & 0.1546 & 0.0549 \\
    & SpatPlus20 & & 0.1852 & 0.0569 & 0.1824 & 0.0521 & 0.1495 & 0.0510 \\
    & SpatPlusP1 & & 0.1141 & 0.0538 & 0.0929 & 0.0506 & 0.0690 & 0.0497 \\
    & SpatPlusTP1 & & 0.1811 & 0.0597 & 0.1546 & 0.0552 & 0.1240 & 0.0541 \\
    & SpatPlusP2 & & 0.0751 & 0.0234 & 0.0692 & 0.0229 & 0.0469 & 0.0234 \\
    & SpatPlusTP2 & & 0.1423 & 0.0392 & 0.1163 & 0.0374 & 0.0775 & 0.0380 \\
    \midrule
    \multirow{12}[2]{*}{\textbf{Scenario 3}} & Null  & \multirow{12}[2]{*}{0.2000} & 0.5706 & 0.0124 & 0.4408 & 0.0131 & 0.3476 & 0.0135 \\
    & Spatial & & 0.4866 & 0.0887 & 0.3718 & 0.0925 & 0.2769 & 0.0964 \\
    & RSR & & 0.5520 & 0.0126 & 0.4232 & 0.0134 & 0.3284 & 0.0139 \\
    & TGRMF1 & & 0.4461 &	0.0792 & 0.3397 & 0.0810 & 0.2450 &	0.0840  \\
    & SpatPlus5 & & 0.3200 & 0.0618 & 0.2426 & 0.0613 & 0.1874 & 0.0620 \\
    & SpatPlus10 & & 0.2020 & 0.0715 & 0.1953 & 0.0643 & 0.1593 & 0.0629 \\
    & SpatPlus15 & & 0.1545 & 0.0635 & 0.1300 & 0.0591 & 0.0989 & 0.0575 \\
    & SpatPlus20 & & 0.1220 & 0.0566 & 0.1215 & 0.0531 & 0.0998 & 0.0525 \\
    & SpatPlusP1 & & 0.0871 & 0.0525 & 0.0579 & 0.0501 & 0.0403 & 0.0494 \\
    & SpatPlusTP1 & & 0.1086 & 0.0608 & 0.1107 & 0.0580 & 0.0913 & 0.0565 \\
    & SpatPlusP2 & & 0.0890 & 0.0169 & 0.1003 & 0.0179 & 0.0815 & 0.0185 \\
    & SpatPlusTP2 & & 0.1526 & 0.0261 & 0.1647 & 0.0281 & 0.1288 & 0.0295 \\
    \bottomrule
    \end{tabular}}
  \label{tab:beta1_simu1}%
\end{table}

Table \ref{tab:beta1_simu1} provides the average over the 100 simulated data sets of the posterior means and posterior standard deviations of the regression coefficient $\beta_1$ obtained with the different models in each simulated Scenario. The results are interesting. In Scenario 1, we observe a highly biased fixed effect estimates for the null, the spatial, the RSR and the TGMRF methods when the correlation between the observed (${\boldsymbol{X}_1}$) and the unobserved (${\boldsymbol{X}_2}$) covariates is high. In this situation, it appears that the estimated $\beta_1$ captures the effect of both covariates $\X_1$ and $\X_2$. The bias reduces when the correlation between the two covariates decreases. In Scenario 1, the spatial+ method with 15 eigenvectors recovers pretty well the true value of $\beta_1$ if the correlation is high. With moderate correlation, 5 or 10 eigenvectors give nearly unbiased estimates. When the correlation is low, the null, the spatial, the RSR, and the TGMRF lead to fixed effects estimates with the lowest bias. Additionally, we observe that the spatial model leads to the highest posterior standard deviation of the fixed effects, whereas the null and RSR models provide the lowest posterior standard deviation. The rest of models provide posterior standard deviations somewhere in between. Results for Scenarios 2 and 3 are somewhat different as additional variability is included through an ICAR model and P-splines respectively. In both scenarios, the null, the spatial, the RSR, and the TGMRF models lead to highly biased fixed effects estimates irrespective of the correlation between  ${\boldsymbol{X}_1}$ and ${\boldsymbol{X}_2}$, though the bias reduces when the correlation decreases. In Scenario 2, the spatial+ methods again recover pretty well the $\beta_1$ coefficient, though now we need to increase the number of eigenvectors in the covariate model. The number of eigenvectors needed is smaller when the correlation between the covariates is low. In this scenario, the TGMRF model produces the highest posterior standard deviations. Similar results are observed in Scenario 3. Here, the highest posterior standard deviations correspond to the spatial model whereas the smallest come from the null and the RSR. In this scenario, the posterior standard errors obtained with the TGMRF models are pretty similar to those of the spatial model.

\begin{figure}[h]
\centering
\includegraphics[width=1\textwidth]{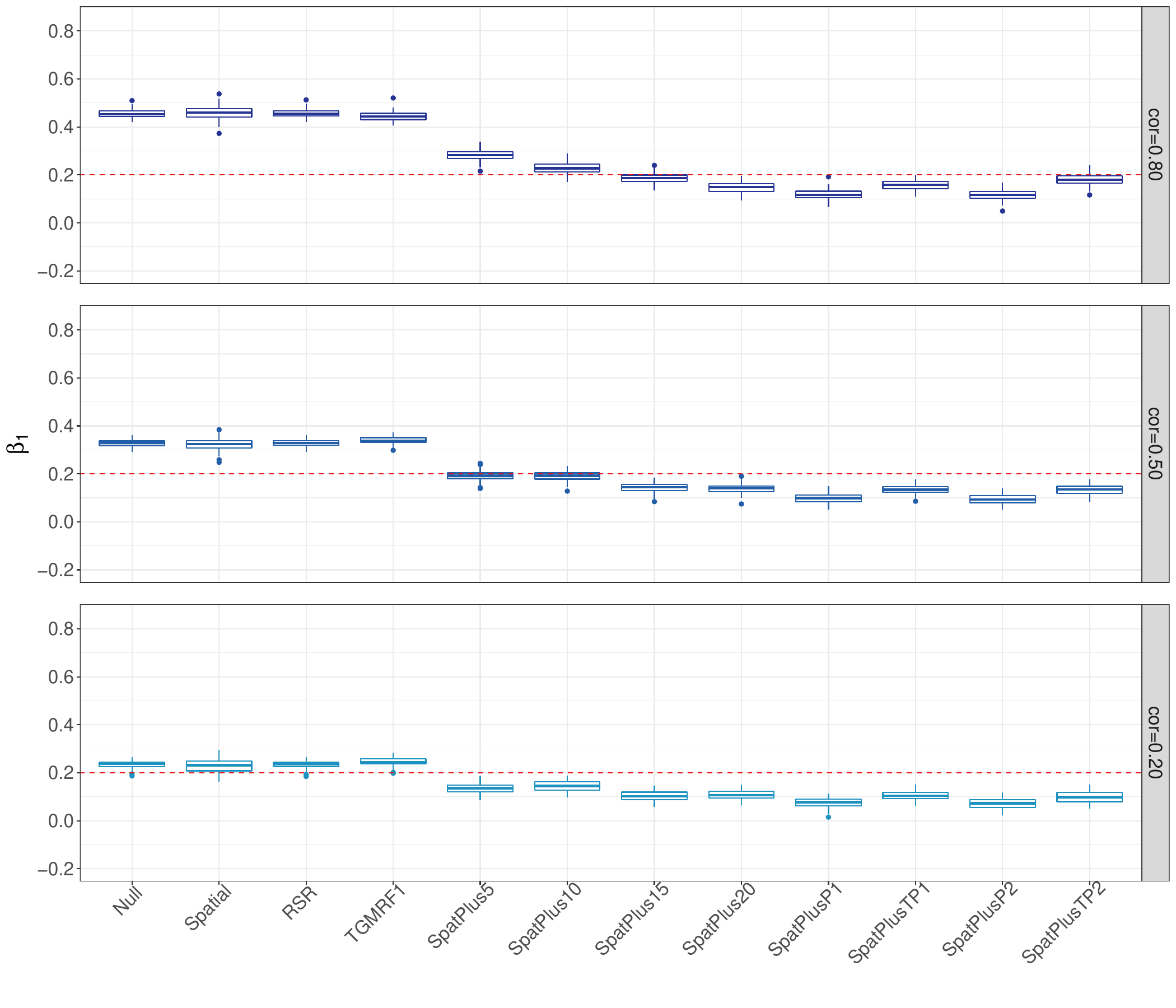}
\caption{Boxplots of the estimated means of $\beta_1$ based on 100 simulated datasets for Simulation study 1,  Scenario 1 and $cor({\X}_1,{\X}_2) = 0.8$ (top row), $0.5$ (middle row), and $0.2$ (bottom row).}
\label{fig:boxplot_scenario1_simu1}
\end{figure}

To inspect visually the different methods, Figure \ref{fig:boxplot_scenario1_simu1} shows the boxplots of the posterior means of $\beta_1$ over the 100 simulated data sets for Scenario 1. The first row shows the boxplots when the correlation between $\X_1$ and $\X_2$ is $0.8$. The second row shows the boxplots when correlation is $0.5$ and the third row shows the boxplots for correlation $0.2$. Figures \ref{fig:boxplot_scenario2_simu1_beta1} and \ref{fig:boxplot_scenario3_simu1_beta1} in the Appendix A display the same boxplots for Scenarios 2 and 3 respectively. Interestingly, the bias of the null, RSR, spatial and TGMRF models reduces when the correlation between the covariates decreases. This reduction is particularly remarkable in Scenario 1. Additionally, Table \ref{tab:MARB_beta1} in the Appendix A provides mean absolute relative bias (MARB) and mean root relative mean squared error (MRRMSE) of the fixed effect estimates to complement the information. For the null, the spatial, the RSR and the TGMRF models both the MARB and the MRRMSE reduce when the correlation between the covariates decreases. This is expected because spatial confounding is more severe if the unobserved covariate is correlated with the observed one. For the rest of models there is not a clear pattern. In general, when the correlation between ${\X}_1$ and ${\X}_2$ is small a spatial+ model with a small number of eigenvectors provides the lowest MARB and MRRMSE. If the correlation is high, a spatial+ model with a larger number of eigenvectors is better.

Table \ref{tab:beta1_simu1} (and Figures \ref{fig:boxplot_scenario1_simu1}, \ref{fig:boxplot_scenario2_simu1_beta1} and \ref{fig:boxplot_scenario3_simu1_beta1}) gives an idea about the magnitude of the bias of the fixed effect estimate as we can compare the average of the posterior means with the true value of $\beta_1$, but they do not give information about the posterior standard deviation. To see if the posterior standard deviation is a good measure of the variability of the fixed effect estimate, Table \ref{tab:simuest} compare the true simulated standard error ($s.e._{sim}$) with the estimated standard error ($s.e._{est}$). They are defined as follows

$$ s.e._{sim}=\sqrt{\frac{1}{100} \sum_{k=1}^{100}\left(\hat{\beta}_1^k-\overline{\hat{\beta}}_1\right)^2}\quad \quad s.e._{est}=\frac{1}{100} \sum_{k=1}^{100} sd(\hat{\beta}_1^k)
$$
where $\hat{\beta}_1^k$ is the posterior mean of $\beta_1$ in simulation $k$, $\overline{\hat{\beta}}_1$ is the average of all the posterior estimates, and $sd(\hat{\beta}_1^k)$ is the posterior standard deviation of $\beta_1$ in simulation $k$. Then, the true simulated standard error is the sample standard deviation of the posterior mean estimates, and the estimated standard error is the average of the posterior standard deviations. If the estimated standard error is higher than the simulated standard error, then we are overestimating the posterior standard deviation of the fixed effects. And the other way around, if the estimated standard error is lower than the simulated standard error we are underestimating the posterior standard deviation of the fixed effects. According to Table \ref{tab:simuest}, the null and the RSR models provides estimated standard errors pretty similar to the simulated ones in all scenarios. On the contrary, the spatial and the TGMRF models lead to estimated standard errors much higher than the simulated ones in all the scenarios. All the spatial+ models tend to overestimate the posterior standard deviation but to a lower extent than the spatial and the TGMRF model. It is worth noting that the spatial+ models SpatPlusP2 and SpatPlusTP2 give pretty similar values of estimated and simulated standard errors.

\begin{table}[htbp]
  \centering
  \caption{Estimated standard errors ($s.e._{est}$) and simulated standard errors ($s.e._{sim}$) for $\beta_1$ based on 100 simulated datasets for Simulation Study 1, Scenarios 1, 2 and 3 and $cor(\X_1,\X_2) = 0.8$, $0.5$ and $0.2$.}
  \resizebox{\textwidth}{!}{
    \begin{tabular}{lccccccc}
    \hline\\
    & & \multicolumn{2}{c}{\textbf{cor=0.80}} & \multicolumn{2}{c}{\textbf{cor=0.50}} & \multicolumn{2}{c}{\textbf{cor=0.20}} \\
    \midrule
    & \textbf{Model} & \multicolumn{1}{c}{$\boldsymbol{s.e._{est}}$} & \multicolumn{1}{c}{$\boldsymbol{s.e._{sim}}$} & \multicolumn{1}{c}{$\boldsymbol{s.e._{est}}$} & \multicolumn{1}{c}{$\boldsymbol{s.e._{sim}}$} & \multicolumn{1}{c}{$\boldsymbol{s.e._{est}}$} & \multicolumn{1}{c}{$\boldsymbol{s.e._{sim}}$} \\
    \midrule
    \multirow{12}[2]{*}{\textbf{Scenario 1}} & Null  & 0.0160 & 0.0169 & 0.0168 & 0.0140 & 0.0175 & 0.0145 \\
    & Spatial & 0.0448 & 0.0262 & 0.0593 & 0.0252 & 0.0665 & 0.0295 \\
    & RSR & 0.0160 & 0.0169 & 0.0171 & 0.0143 & 0.0178 & 0.0146 \\
    & TGRMF1 & 0.0313 & 0.0192 & 0.0449 & 0.0151 & 0.0505 & 0.0169 \\
    & SpatPlus5 & 0.0377 & 0.0215 & 0.0407 & 0.0194 & 0.0429 & 0.0209 \\
    & SpatPlus10 & 0.0438 & 0.0238 & 0.0398 & 0.0205 & 0.0403 & 0.0214 \\
    & SpatPlus15 & 0.0410 & 0.0221 & 0.0386 & 0.0198 & 0.0383 & 0.0206 \\
    & SpatPlus20 & 0.0401 & 0.0209 & 0.0361 & 0.0199 & 0.0359 & 0.0200 \\
    & SplatPlusP1 & 0.0386 & 0.0209 & 0.0361 & 0.0202 & 0.0356 & 0.0206 \\
    & SpatPlusTP1 & 0.0409 & 0.0204 & 0.0378 & 0.0187 & 0.0376 & 0.0191 \\
    & SpatPlusP2 & 0.0206 & 0.0193 & 0.0213 & 0.0203 & 0.0218 & 0.0213 \\
    & SpatPlusTP2 & 0.0261 & 0.0224 & 0.0267 & 0.0214 & 0.0281 & 0.0236 \\
    \midrule
    \multirow{12}[2]{*}{\textbf{Scenario 2}} & Null  & 0.0144 & 0.0112 & 0.0149 & 0.0130 & 0.0154 & 0.0138 \\
    & Spatial & 0.0834 & 0.0307 & 0.0900 & 0.0264 & 0.0949 & 0.0292 \\
    & RSR & 0.0152 & 0.0129 & 0.0160 & 0.0153 & 0.0165 & 0.0164 \\
    & TGRMF1 & 0.0840 &	0.0306 & 0.0951 & 0.0339 & 0.1001 &	0.0321 \\
    & SpatPlus5 & 0.0627 & 0.0251 & 0.0627 & 0.0202 & 0.0630 & 0.0212 \\
    & SpatPlus10 & 0.0699 & 0.0279 & 0.0642 & 0.0248 & 0.0625 & 0.0242 \\
    & SpatPlus15 & 0.0603 & 0.0233 & 0.0561 & 0.0229 & 0.0549 & 0.0222 \\
    & SpatPlus20 & 0.0569 & 0.0234 & 0.0521 & 0.0213 & 0.0510 & 0.0211 \\
    & SpatPlusP1 & 0.0538 & 0.0251 & 0.0506 & 0.0245 & 0.0497 & 0.0247 \\
    & SpatPlusTP1 & 0.0597 & 0.0218 & 0.0552 & 0.0228 & 0.0541 & 0.0227 \\
    & SpatPlusP2 & 0.0234 & 0.0202 & 0.0229 & 0.0217 & 0.0234 & 0.0224 \\
    & SpatPlusTP2 & 0.0392 & 0.0330 & 0.0374 & 0.0320 & 0.0380 & 0.0330 \\
    \midrule
    \multirow{12}[2]{*}{\textbf{Scenario 3}} & Null  & 0.0124 & 0.0124 & 0.0131 & 0.0116 & 0.0135 & 0.0136 \\
    & Spatial & 0.0887 & 0.0243 & 0.0925 & 0.0227 & 0.0964 & 0.0243 \\
    & RSR & 0.0126 & 0.0138 & 0.0134 & 0.0134 & 0.0139 & 0.0155 \\
    & TGRMF1 & 0.0792 &	0.0217 & 0.0810 & 0.0204 & 0.0840 &	0.0215 \\
    & SpatPlus5 & 0.0618 & 0.0176 & 0.0613 & 0.0169 & 0.0620 & 0.0168 \\
    & SpatPlus10 & 0.0715 & 0.0236 & 0.0643 & 0.0241 & 0.0629 & 0.0209 \\
    & SpatPlus15 & 0.0635 & 0.0200 & 0.0591 & 0.0186 & 0.0575 & 0.0185 \\
    & SpatPlus20 & 0.0566 & 0.0166 & 0.0531 & 0.0164 & 0.0525 & 0.0168 \\
    & SpatPlusP1 & 0.0525 & 0.0186 & 0.0501 & 0.0222 & 0.0494 & 0.0218 \\
    & SpatPlusTP1 & 0.0608 & 0.0200 & 0.0580 & 0.0191 & 0.0565 & 0.0196 \\
    & SpatPlusP2 & 0.0169 & 0.0157 & 0.0179 & 0.0203 & 0.0185 & 0.0176 \\
    & SpatPlusTP2 & 0.0261 & 0.0224 & 0.0281 & 0.0298 & 0.0295 & 0.0253 \\
    \bottomrule
    \end{tabular}}
  \label{tab:simuest}%
\end{table}%

In addition to the posterior mean and standard deviation, and to have a complete view on the inference about fixed effects, we are interested in credible intervals. Table \ref{tab:coverage_simu1} displays the empirical coverage of credible intervals for $\beta_1$ at $95\%$ nominal value.
In general, the empirical coverage obtained with the null and the RSR models is very low, in many cases 0. This is expected because of the high bias. Regarding the spatial model, the empirical coverage is also very low. Again this is explained by the high bias. However, in Scenario 1 and 3 when correlation is 0.2 the coverage is 100\% and in Scenario 2 and $cor({\boldsymbol X}_1,{\boldsymbol X}_2)=0.2$ the coverage is 92\%, close to the nominal value. The performance of the TGMRF is similar to the spatial model. Regarding the spatial+ method using eigenvectors of the precision matrix, we observe in general overcoverage. This can be explained because the method reduces the bias but overestimates the standard error. In some cases we observe a clear under-coverage that is explained because the overestimation of the standard error does not compensate for the bias. In general, the over-coverage is due to large standard errors whereas under-coverage can be attributed to a large bias. To have a complete picture about coverages, Table \ref{tab:length_CI} in the Appendix A provides the length of the 95\% credible intervals for the parameter $\beta_1$ obtained with the different methods. The most remarkable point is that the null and RSR models give substantially shorter credible intervals than the other models. The widest credible intervals are obtained with the spatial and the TGMRF models, and the spatial+ models give credible intervals wider than the null and RSR models but narrower than the spatial and the TGMRF models.

\begin{table}[htbp]
  \centering
  \caption{Empirical $95\%$ coverage probabilities of the true value of $\beta_1$ based on 100 simulated datasets for Scenarios 1, 2, and 3 and $cor({\X}_1, {\X}_2)=0.8, 0.5$ and $0.2$.}
  \resizebox{\textwidth}{!}{
    \begin{tabular}{lccccccccc}
    \hline\\
     & \multicolumn{3}{c}{\textbf{Scenario 1}} & \multicolumn{3}{c}{\textbf{Scenario 2}} & \multicolumn{3}{c}{\textbf{Scenario 3}} \\
\cmidrule{1-10}     \textbf{Model}    & \textbf{cor=0.80} & \textbf{cor=0.50} & \textbf{cor=0.20} & \textbf{cor=0.80} & \textbf{cor=0.50} & \textbf{cor=0.20} & \textbf{cor=0.80} & \textbf{cor=0.50} & \textbf{cor=0.20} \\
    \midrule
    Null  & 0     & 0     & 44    & 0     & 0     & 0     & 0     & 0     & 0 \\
    Spatial  & 0     & 40    & 100   & 0     & 1     & 92    & 0     & 70    & 100 \\
    RSR & 0     & 0     & 49    & 0     & 0     & 0     & 0     & 0     & 0 \\
    TGMRF1   & 0 & 0 & 100 & 0 & 0 & 44 & 0	& 71  & 100 \\
    SpatPlus5 & 33    & 100   & 80    & 18    & 100   & 100   & 49    & 100   & 100 \\
    SpatPlus10 & 100   & 100   & 82    & 100   & 100   & 100   & 100   & 100   & 100 \\
    SpatPlus15 & 100   & 84    & 23    & 100   & 100   & 100   & 100   & 100   & 69 \\
    SpatPlus20 & 89    & 73    & 17    & 100   & 100   & 100   & 99    & 93    & 62 \\
    SpatPlusP1 & 37    & 4     & 0     & 79    & 34    & 9     & 30    & 2     & 0 \\
    SpatPlusTP1 & 95    & 71    & 20    & 100   & 100   & 92    & 90    & 89    & 59 \\
    SpatPlusP2 & 1     & 0     & 0     & 0     & 0     & 0     & 0     & 0     & 0 \\
    SpatPlusTP2 & 92    & 30    & 4     & 68    & 31    & 8     & 54    & 72    & 33 \\
    \bottomrule
    \end{tabular}}
  \label{tab:coverage_simu1}
\end{table}


To complete this simulation study, we would like to have a look at risk smoothing and goodness of fit. Table \ref{tab:DIC_simu} in the Appendix A displays averages over the 100 simulations of  WAIC values. The null model is clearly insufficient for risk smoothing and presents larger WAIC values than the other methods. The spatial+ models SpatPlusP2 and SpatPlusTP2 also provide larger values of WAIC than the other models. Probably they are oversmoothing the risks. Differences among the rest of models are minor indicating that the procedures lead to a similar smoothing. This is corroborated in Table \ref{tab:MARB_risk} of the Appendix A where MARB and MRRMSE of the relative risks are provided. In general, the null model and the spatial+ models SpatPlusP2 and SpatPlusTP2 give the largest MARB and MRRMSE, indicating a worse fit. The rest of models provide MARBs below 10\%.

Finally, as suggested by one reviewer, we have simulated a Scenario 4 where the additional term ${\boldsymbol S}$ has been generated from a multivariate normal distributions $N({\boldsymbol 0},\sigma^2{\boldsymbol I}_n)$ with $\sigma^2=0.2$. That is, the additional variability is not spatially structured. Results are rather similar to those from Scenario 3 and they are not shown to save space. The reason why the results are similar is probably because the correlation between the generated random effects and the covariate ${\X}_1$ in Scenario 4 (0.1438) is very similar to the correlation between the spatial surface and the covariate ${\X}_1$ in Scenario 3 (0.1998).

\subsection{Simulation study 2: Results}

To complete the study, we now pay attention to the Type-S error rate of the different methods considered in this paper.

Table \ref{tab:typeS_error_simu2} displays the Type-S errors for $\beta_2$ based on 100 simulated datasets for each scenario. Type-S error rates should be around the nominal value 5\%. In Scenario 1, where there is no more variability than that introduced by the covariate $\boldsymbol{X}_1$, the Type-S error rate is small (less than 10\%) for all the methods. This agrees with the results of \cite{Khan2020}. In Scenarios 2 and 3, where additional variability is introduced in the generating model through an ICAR and P-splines respectively, the Type-S error rates are very high for the null and the RSR model. This is in line with some results in \cite{Hanks2015}. Overall, the spatial+ models do not produce high Type-S error rates. The exception is Scenario 2 and high correlation between the covariates where the models SpatPlusP2 and SpatPlusTP2 exhibit rates over 30\%. To better understand the Type-S error rates in Table \ref{tab:typeS_error_simu2}, Figures \ref{fig:boxplot_scenario1_simu2_beta2}, \ref{fig:boxplot_scenario2_simu2_beta2}, and \ref{fig:boxplot_scenario3_simu2_beta2} in the Appendix A display the posterior mean estimates of the parameter $\beta_2$. The bias of the null and the RSR models in Scenarios 2 and 3 helps to understand the high Type-S error rates in some subscenarios.

\begin{table}[htbp]
  \centering
  \caption{Type-S errors rate (\%) of $\beta_2$ based on 100 simulated datasets for Scenarios 1, 2 and 3 and $cor({\X}_1, {\X}_2)=0.8, 0.5$ and $0.2$.}
  \resizebox{\textwidth}{!}{
    \begin{tabular}{lccccccccc}
    \hline\\
          & \multicolumn{3}{c}{\textbf{Scenario 1}} & \multicolumn{3}{c}{\textbf{Scenario 2}} & \multicolumn{3}{c}{\textbf{Scenario 3}} \\
\cmidrule{1-10}  \textbf{Model} & \textbf{cor=0.80} & \textbf{cor=0.50} & \textbf{cor=0.20} & \textbf{cor=0.80} & \textbf{cor=0.50} & \textbf{cor=0.20} & \textbf{cor=0.80} & \textbf{cor=0.50} & \textbf{cor=0.20} \\
    \midrule
    Null  & 9     & 2     & 2     & 8     & 95    & 95    & 98    & 12    & 12 \\
    Spatial  & 8     & 2     & 2     & 11    & 0     & 0     & 0     & 0     & 0 \\
    RSR & 9     & 2     & 2     & 13    & 96    & 96    & 98    & 24    & 24 \\
    TGMRF1   &  6 & 1 & 1 & 1 & 0 & 0 & 0 & 0 & 0\\
    SpatPlus5 & 2     & 3     & 3     & 12    & 0     & 0     & 0     & 0     & 0 \\
    SpatPlus10 & 3     & 1     & 1     & 11    & 0     & 0     & 0     & 0     & 0 \\
    SpatPlus15 & 2     & 0     & 0     & 12    & 0     & 0     & 0     & 0     & 0 \\
    SpatPlus20 & 2     & 0     & 0     & 3     & 0     & 0     & 0     & 0     & 0 \\
    SpatPlusP1 & 5     & 6     & 1     & 15    & 0     & 0     & 0     & 0     & 0 \\
    SpatPlusTP1 & 3     & 0     & 0     & 4     & 0     & 0     & 0     & 0     & 0 \\
    SpatPlusP2 & 9     & 5     & 4     & 36    & 7     & 5     & 8     & 2     & 4 \\
    SpatPlusTP2 & 6     & 2     & 2     & 31    & 6     & 6     & 6     & 5     & 5 \\
    \bottomrule
    \end{tabular}}
  \label{tab:typeS_error_simu2}
\end{table}%


\section{Discussion}

Spatial confounding is a problem that still remains unsolved or at least partially unsolved. One of the main difficulties is that there is not a unique and general definition. Traditionally, spatial confounding has been considered as a collinearity problem between the fixed and the random effects. Or in other words, the fixed and random effects \lq\lq compete'' for the same variability. Then, when random effects with a spatial correlation structure are included in a linear or generalized linear model, the fixed effects estimates change. The question is if we should expect a change or not.

One of the most popular methods to deal with spatial confounding, restricted spatial regression, was proposed to avoid the change in fixed effects estimates in relation to the model without spatial random effects. Restricted spatial regression simply restricts the random effects to lie in the orthogonal complement of the fixed effects, consequently the fixed effects estimates do not change. The idea underlying restricted regression is to assign all the variability in the direction of the covariates to the covariates themselves. This seems a good idea if we assume that the estimates we obtain in the null model (the one without spatial random effects) are correct. If this is not the case, and a spatial random effect is introduced in a model to deal with the remaining spatial variability that the observed covariates do not account for, some issues arise. The main one is collinearity, because the spatial random effects also compete to explain the same variability as the observed covariates. Restricted spatial regression implicitly assumes that there are no other covariates overlapping with the observed ones, something that might not be very realistic in practice. On the other hand, the standard errors of the fixed effects estimates in the null model is known to be too small and they are inflated when the spatial random effect is included in the model. The restricted regression was supposed to provide standard errors for the fixed effects estimates somewhere in between. However, recent research \citep[see for example][]{Khan2020,Zimmerman2021} shows that with normal responses, the restricted regression provides standard errors less than or equal to those obtained with the null model. Consequently, inference is liberal and Type-S error rates can be high. However, with Poisson responses, no clear results have been provided yet.
In this line, and assuming that spatial random effects play the role of unobserved covariates with spatial structure, recent research \citep{Gilbert2021} suggests that a change in the fixed effects estimates is expected and collinearity between fixed and random effects is not a problem because this collinearity represents the overlap between observed and unobserved covariates. These authors study spatial confounding from a causal inference perspective, where the change in the fixed effect estimates is due to the existence of unmeasured variables spatially structured.


Given the controversy about spatial confounding, in this paper we analyse three data sets to illustrate how different techniques yield to different estimates and posterior standard deviations and hence, produce different conclusions about the fixed effects. Then, we run a simulation study to evaluate how some of the different existing methods designed to alleviate spatial confounding estimate the fixed effects in different scenarios. Namely, a simple Poisson regression model, a Poisson spatial mixed model, restricted  spatial regression, TGMRFs and spatial+ models. Spatial confounding is introduced by using generating models with two covariates, ${\X}_1$ and ${\X}_2$, where the first one plays the role of the observed covariate and the second one acts as an unobserved covariate that is not included in the fitting process. Additional spatial variability is added in the generating process using an ICAR spatial random effect or a spatial surface generated using P-splines. More precisely, in Scenario 1 all the variability is introduced with the covariates. In Scenario 2, additional spatial variability is included with an ICAR random effect, and finally, in Scenario 3, we use a spatial surface generated using P-splines to introduced additional spatial variability in the generating process. The results of the simulation study are very informative. Overall, the method that best recovers the true value of the fixed effects is the spatial+ model using eigenvectors of the spatial precision matrix as regressors in the covariate model. The number of eigenvectors depends on the correlation between the two covariates ${\X}_1$ and ${\X}_2$, and on the way we generate additional spatial variability (ICAR or P-splines). In general, the higher the correlation between the covariates, the larger the number of eigenvectors. When the correlation is high (0.8), 14-21\% of the eigenvectors associated to the lower eigenvalues of the spatial precision matrix are required. If the correlation is medium (0.5), 7-14\% of the eigenvectors are needed if the generating model only includes the covariates, whereas if the generating model includes additional variability (ICAR or P-splines), 14-21\% of the eigenvectors seem to produce good results. Finally, when the correlation between the covariates is low (0.2), 7-14\% eigenvectors are needed in Scenarios 2 and 3. However,  the spatial+ model does not provide good results in Scenario 1 where there is no additional spatial variability other than that included in the covariates.

In terms of standard errors, the posterior standard deviation in the null and in the RSR models seems to be a good estimator of the true standard error, whereas the rest of the models tend to overestimate the true standard error, notably the spatial and the TGMRF models. Regarding coverage rates, it seems that the spatial+ method leads to overcoverage, something expected as it also overestimates the standard error. In addition, the Type-S error rates are very low in several scenarios. Therefore, the spatial+ method with a suitable number of eigenvectors seems to recover the true fixed effects quite well but could inflate standard errors. In our opinion, Scenarios 2 and 3 are the most realistic as they include additional spatial variability other than that captured by the covariates and a number of eigenvectors between 14\%  and 21\% of the total could be a good choice in general.


Regarding risk estimation, the null model is clearly insufficient, whereas similar estimates are obtained with the rest of models with the exception of the spatial+ using splines (P-splines of thin plate splines) to smooth the risks. This agrees with the work by \cite{Adin2021}, where identical risk estimates where observed with the spatial and the restricted spatial regression models. As suggested by one reviewer, we have also generated a Scenario 4 where the additional variability is spatially unstructured. It is worth noting that results in this scenario are rather similar to those of Scenario 3, so they have been omitted to save space. 
We remark that if researchers are interested in risk prediction, probably the fixed effects estimates are not so important given that all the spatial methods including ICAR random effects lead to essentially identical risk surfaces, i.e., irrespective of the fixed effect estimated value, the risk predictions do not change. However, if researchers are interested in identifying potential risk factors looking at the spatial map of the unexplained variability, it is crucial to provide unbiased estimates of the fixed effects, otherwise the map of the remaining variability would not be correct.

To conclude this paper, we provide some guidelines to practitioners in light of our simulation results. Our advice is to fit the null and the spatial model first. If there is no change in the fixed effects estimates, probably spatial confounding is not an issue. If a substantial change is observed, the spatial+ method with a number of eigenvectors between 14\%  and 21\% of the total could lead to nearly unbiased fixed effects estimates. However, inference could probably be too conservative as the method seems to inflate standard errors.  This might be what we observe in the real data analyses of this paper. In any case, caution is always recommended as our results depend on the generating models, and different data generating mechanisms could lead to different conclusions.


\section*{Declarations}

\subsection*{Funding and Conflicts of interests/Competing interests}
This work has been supported by Project PID2020-113125RB-I00/ MCIN/ AEI/ 10.13039/501100011033.

The authors have no competing interests to declare that are relevant to the content of this article.

\newpage
\bibliographystyle{apalike}
\bibliography{references}

\newpage
\setcounter{section}{0} 
\renewcommand{\thesection}{\Alph{section}}

\setcounter{figure}{0}
\renewcommand\thefigure{\thesection.\arabic{figure}}

\setcounter{table}{0}
\renewcommand\thetable{\thesection.\arabic{table}}

\section{Supplementary material}

This Supplementary Material contains the following tables and figures to complement the paper \lq\lq Evaluating recent methods to overcome spatial confounding".

\begin{figure}[h]
\centering
\includegraphics[width=1\textwidth]{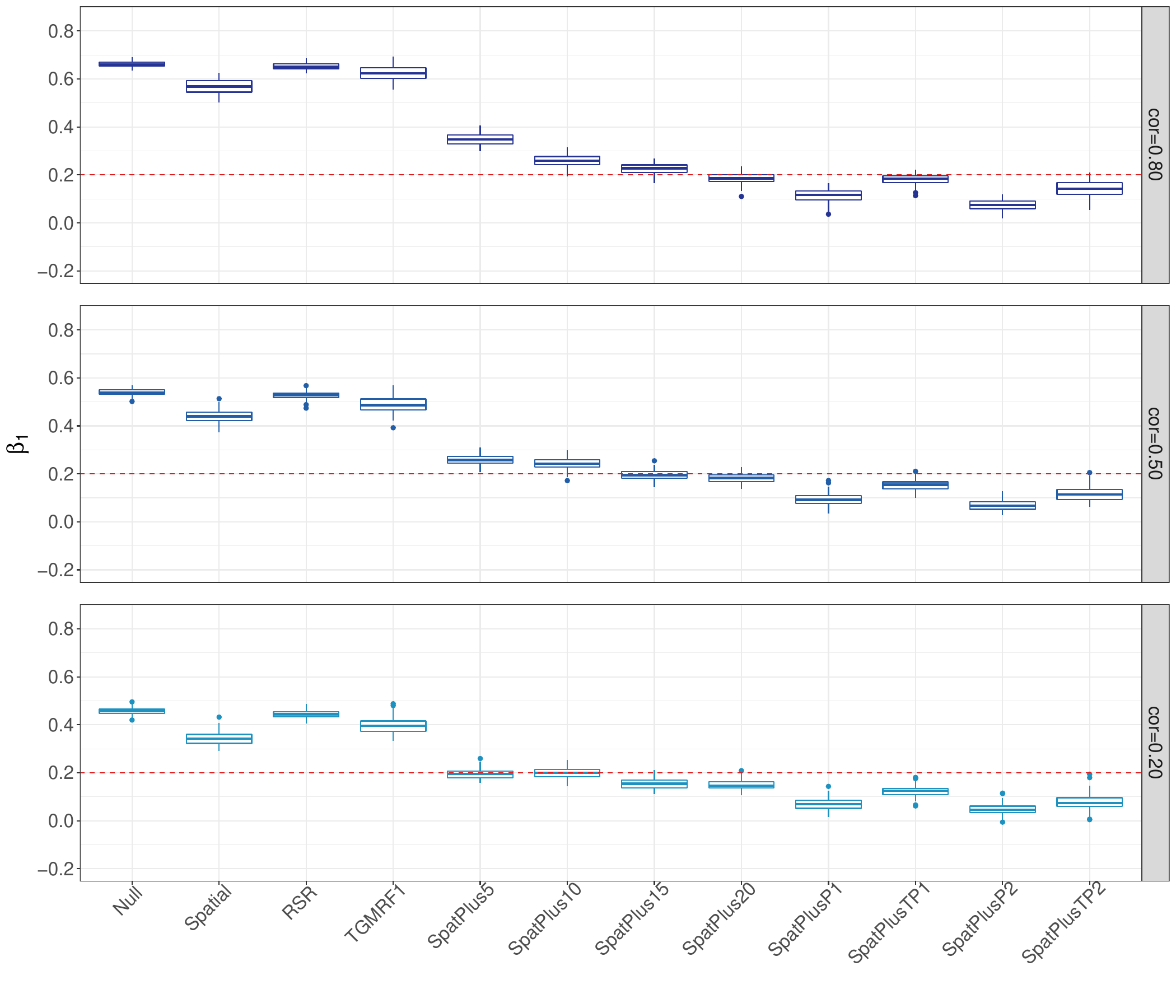}
\caption{Boxplots of the estimated means of $\beta_1$ based on 100 simulated datasets for Simulation study 1, Scenario 2 and $cor({\X}_1,{\X}_2) = 0.8$ (top row), $0.5$ (middle row), and $0.2$ (bottom row).}
\label{fig:boxplot_scenario2_simu1_beta1}
\end{figure}

\begin{figure}[h]
\centering
\includegraphics[width=1\textwidth]{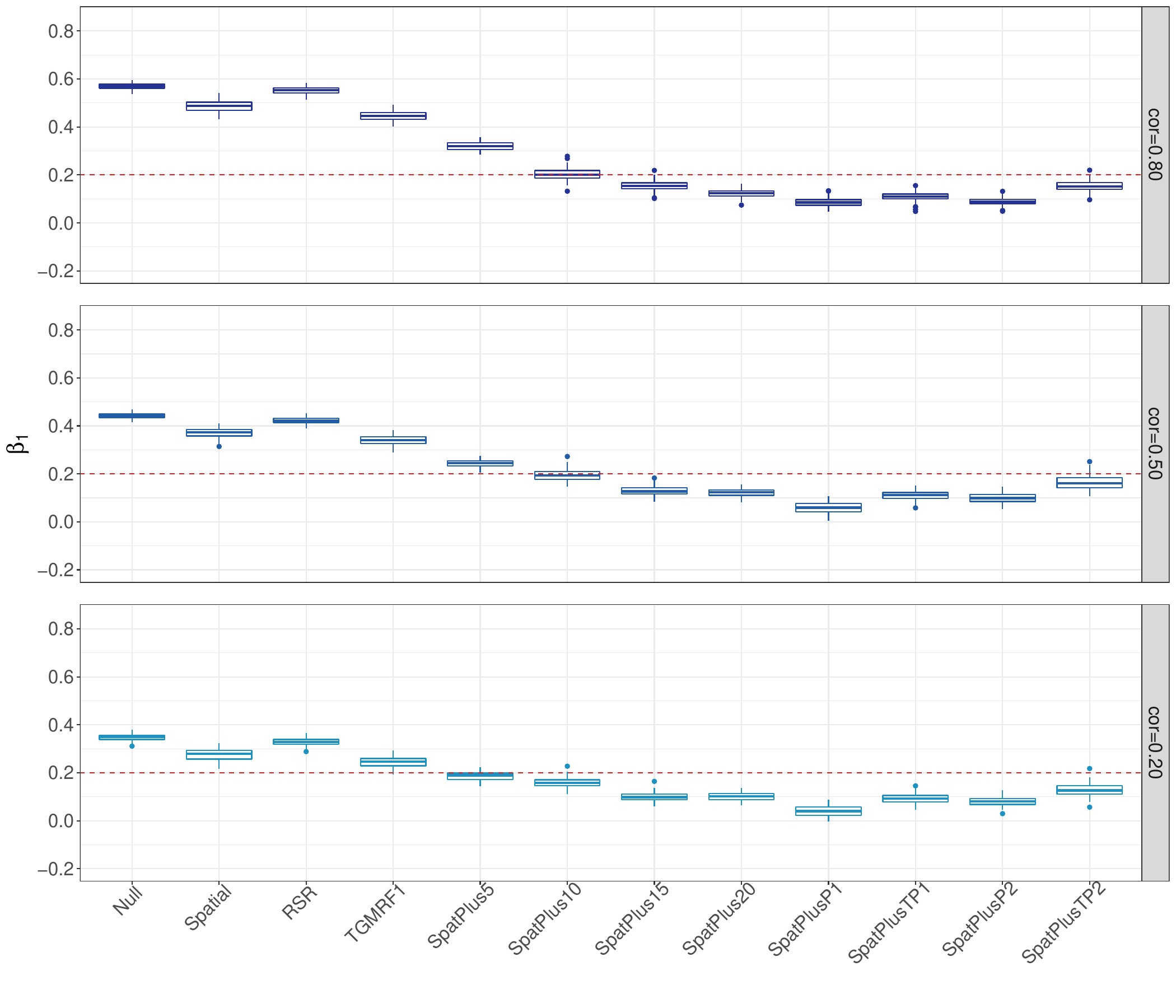}
\caption{Boxplots of the estimated means of $\beta_1$ based on 100 simulated datasets for Simulation study 1,  Scenario 3 and $cor({\X}_1,{\X}_2) = 0.8$ (top row), $0.5$ (middle row), and $0.2$ (bottom row).}
\label{fig:boxplot_scenario3_simu1_beta1}
\end{figure}

\begin{figure}[h]
\centering
\includegraphics[width=1\textwidth]{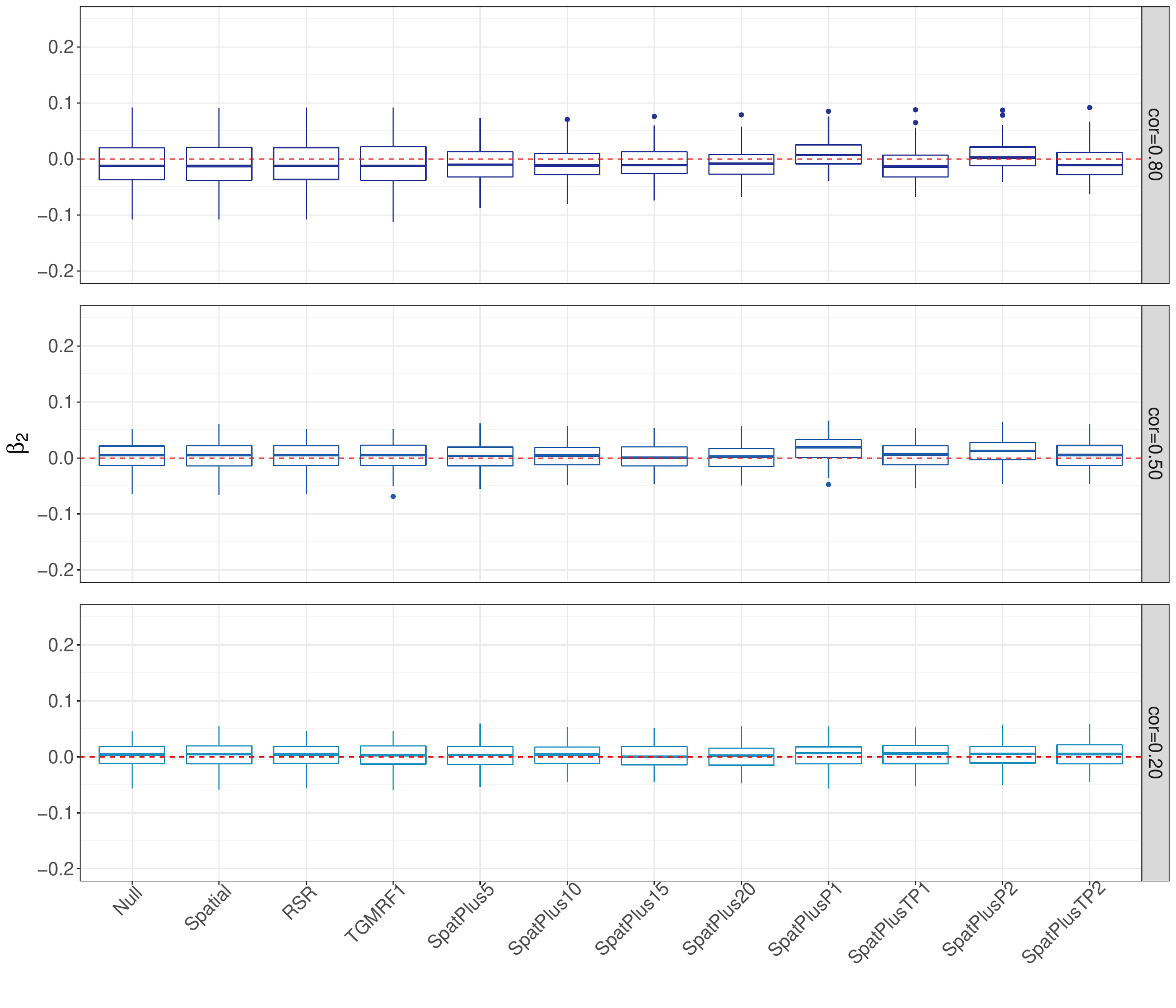}
\caption{Boxplots of the estimated means of $\beta_2$ based on 100 simulated datasets for Simulation study 2, Scenario 1 and $cor({\X}_1, {\X}_2) = 0.8$ (top row), $0.5$ (middle row), and $0.2$ (bottom row).}
\label{fig:boxplot_scenario1_simu2_beta2}
\end{figure}

\begin{figure}[h]
\centering
\includegraphics[width=1\textwidth]{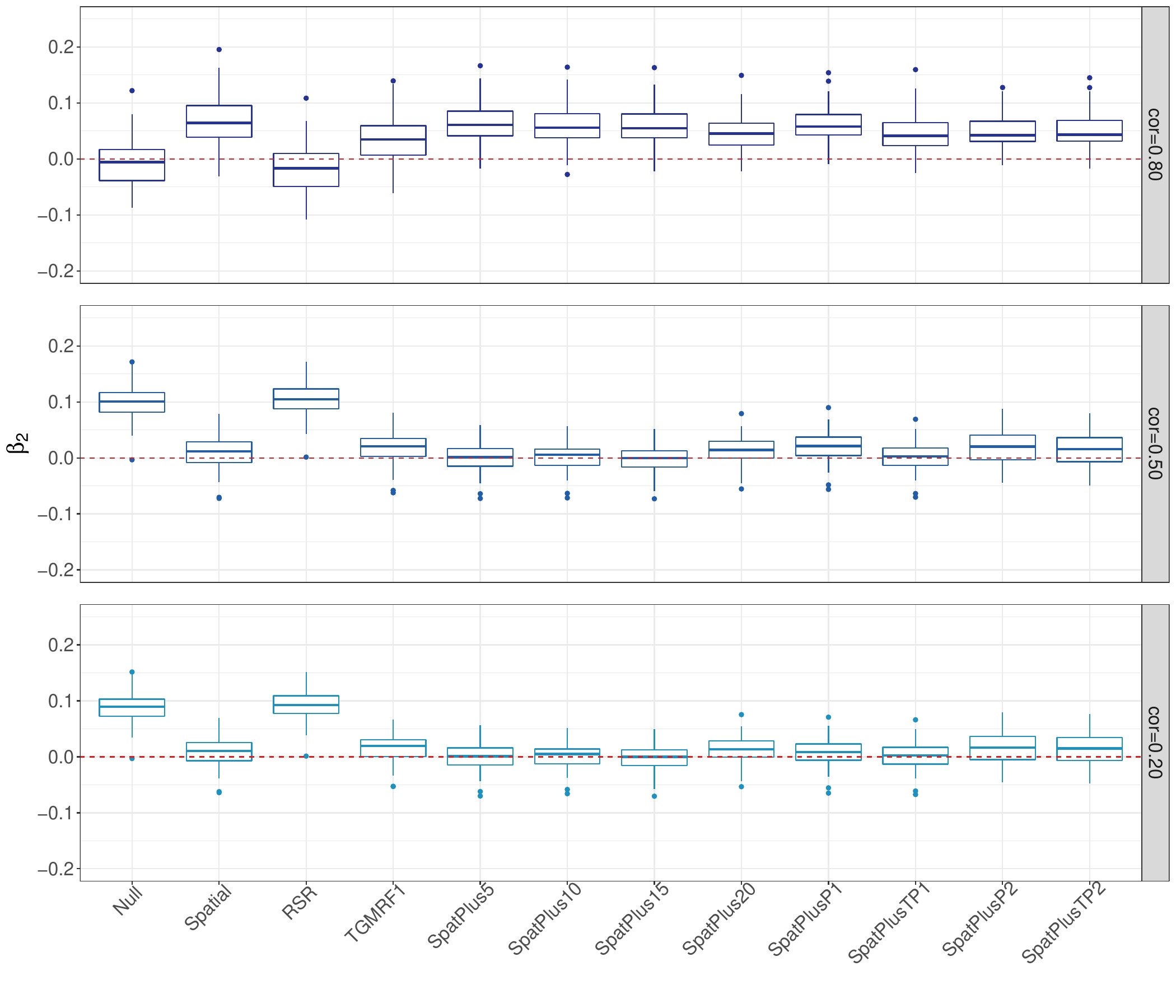}
\caption{Boxplots of the estimated means of $\beta_2$ based on 100 simulated datasets for Simulation study 2, Scenario 2 and $cor({\X}_1, {\X}_2) = 0.8$ (top row), $0.5$ (middle row), and $0.2$ (bottom row).}
\label{fig:boxplot_scenario2_simu2_beta2}
\end{figure}

\begin{figure}[h]
\centering
\includegraphics[width=1\textwidth]{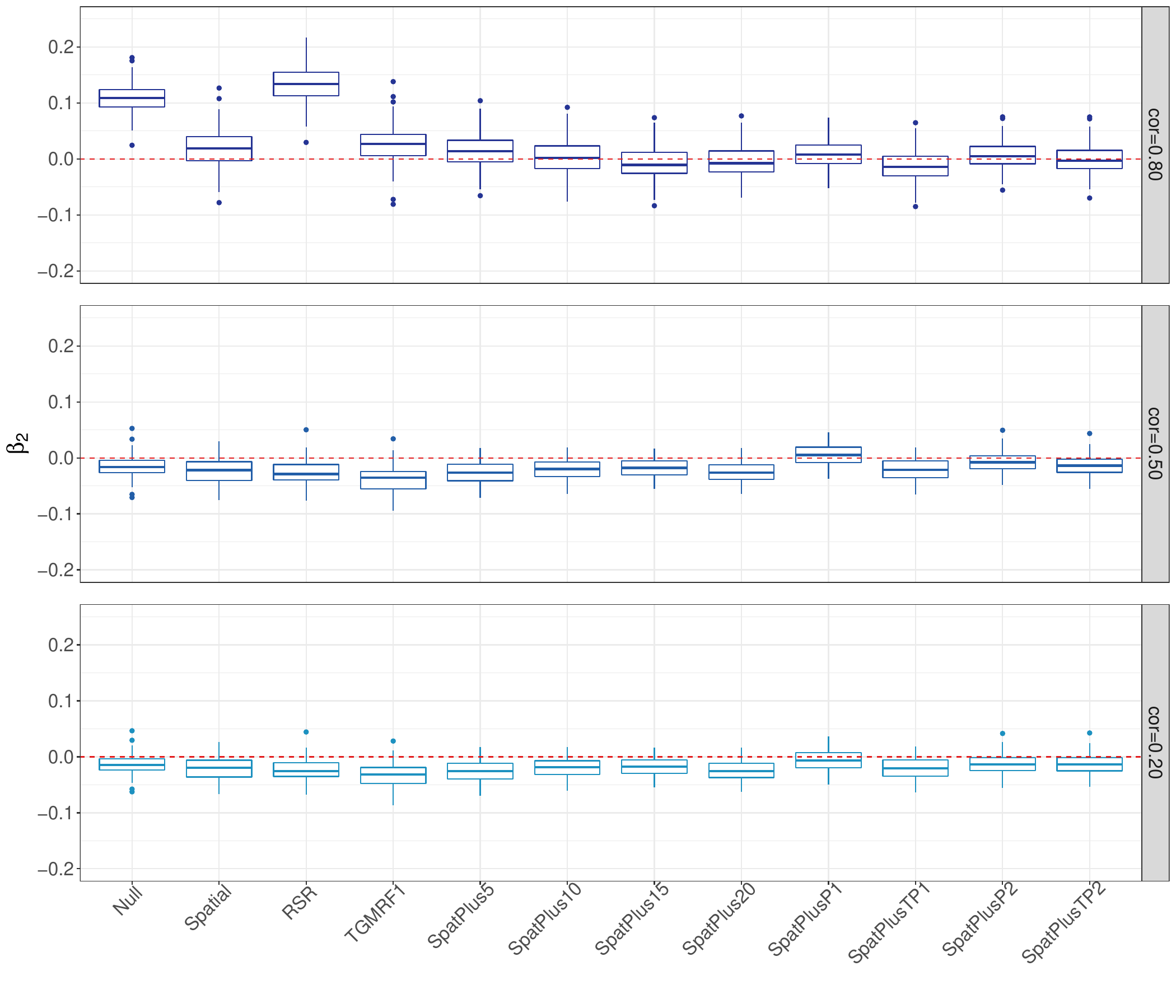}
\caption{Boxplots of the estimated means of $\beta_2$ based on 100 simulated datasets for Simulation study 2, Scenario 3 and $cor({\X}_1, {\X}_2) = 0.8$ (top row), $0.5$ (middle row), and $0.2$ (bottom row).}
\label{fig:boxplot_scenario3_simu2_beta2}
\end{figure}

\begin{enumerate}
  \item Figure \ref{fig:boxplot_scenario2_simu1_beta1}: Boxplots of the estimated means of $\beta_1$ based on 100 simulated datasets for Simulation study 1,  Scenario 2 and $cor({\X}_1,{\X}_2) = 0.8$, $0.5$, and $0.2$.
  \item Figure \ref{fig:boxplot_scenario3_simu1_beta1}: Boxplots of the estimated means of $\beta_1$ based on 100 simulated datasets for Simulation study 1,  Scenario 3 and $cor({\X}_1,{\X}_2) = 0.8$, $0.5$, and $0.2$.
  \item Table \ref{tab:MARB_beta1}: Average value of mean absolute relative bias (MARB) and mean relative root mean prediction error (MRRMSE) of $\beta_1$ based on 100 simulated data sets for Simulation Study 1, Scenarios 1, 2 and 3 and $cor(\X_1,\X_2) = 0.8$, $0.5$ and $0.2$.
  \item Table \ref{tab:length_CI}: Length of the $95\%$ credible intervals of $\beta_1$ for Simulation Study 1, Scenarios 1, 2 and 3 and $cor(\X_1,\X_2) = 0.8$, $0.5$ and $0.2$.
  \item Table \ref{tab:DIC_simu}:  WAIC values based on 100 simulated data sets for Simulation Study 1, Scenarios 1, 2 and 3 and $cor(\X_1,\X_2) = 0.8$, $0.5$ and $0.2$.
  \item Table \ref{tab:MARB_risk}: Average value of mean absolute relative bias (MARB) and mean relative root mean prediction error (MRRMSE) of the relative risks based on 100 simulated data sets for Simulation Study 1, Scenarios 1, 2 and 3 and $cor(\X_1,\X_2) = 0.8$, $0.5$ and $0.2$.
  \item Figure \ref{fig:boxplot_scenario1_simu2_beta2}: Boxplots of the estimated means of $\beta_2$ based on 100 simulated datasets for Simulation study 2, Scenario 1 and $cor({\X}_1, {\X}_2) = 0.8$, $0.5$, and $0.2$.
    \item Figure \ref{fig:boxplot_scenario2_simu2_beta2}: Boxplots of the estimated means of $\beta_2$ based on 100 simulated datasets for Simulation study 2, Scenario 2 and $cor({\X}_1, {\X}_2) = 0.8$, $0.5$, and $0.2$.
    \item Figure \ref{fig:boxplot_scenario3_simu2_beta2}: Boxplots of the estimated means of $\beta_2$ based on 100 simulated datasets for Simulation study 2, Scenario 3 and $cor({\X}_1, {\X}_2) = 0.8$, $0.5$, and $0.2$.
\end{enumerate}

\begin{table}[h]
  \centering
  \caption{Average value of mean absolute relative bias (MARB) and mean relative root mean prediction error (MRRMSE) of $\beta_1$ based on 100 simulated data sets for Simulation Study 1, Scenarios 1, 2 and 3 and $cor(\X_1,\X_2) = 0.8$, $0.5$ and $0.2$.}
  \resizebox{\textwidth}{!}{
  \begin{tabular}{lccccccc}
  \hline\\
  & & \multicolumn{2}{c}{\textbf{cor=0.80}} & \multicolumn{2}{c}{\textbf{cor=0.50}} & \multicolumn{2}{c}{\textbf{cor=0.20}} \\
  \midrule
  & \textbf{Model} & \multicolumn{1}{c}{\textbf{MARB}} & \multicolumn{1}{c}{\textbf{MRRMSE}} & \multicolumn{1}{c}{\textbf{MARB}} & \multicolumn{1}{c}{\textbf{MRRMSE}} & \multicolumn{1}{c}{\textbf{MARB}} & \multicolumn{1}{c}{\textbf{MRRMSE}} \\
  \midrule
    \multirow{12}[2]{*}{\textbf{Scenario 1}} & Null  & 1.2755 & 1.2783 & 0.6387 & 0.6425 & 0.1763 & 0.1906 \\
    & Spatial & 1.3014 & 1.3080 & 0.6199 & 0.6326 & 0.1495 & 0.2099 \\
    & RSR & 1.2845 & 1.2873 & 0.6387 & 0.6428 & 0.1714 & 0.1864 \\
    & TGRMF1 & 1.2229 &	1.2266 & 0.6916 & 0.6957 & 0.2301 &	0.2451 \\
    & SpatPlus5 & 0.4188 & 0.4324 & 0.0302 & 0.1017 & 0.3203 & 0.3370 \\
    & SpatPlus10 & 0.1422 & 0.1855 & 0.0399 & 0.1100 & 0.2786 & 0.2985 \\
    & SpatPlus15 & 0.0694 & 0.1306 & 0.2819 & 0.2987 & 0.4857 & 0.4965 \\
    & SpatPlus20 & 0.2572 & 0.2775 & 0.3056 & 0.3214 & 0.4613 & 0.4720 \\
    & SpatPlusP1 & 0.4115 & 0.4245 & 0.5070 & 0.5170 & 0.6186 & 0.6272 \\
    & SpatPlusTP1 & 0.2106 & 0.2340 & 0.3252 & 0.3384 & 0.4728 & 0.4823 \\
    & SpatPlusP2 & 0.4129 & 0.4240 & 0.5278 & 0.5374 & 0.6375 & 0.6463 \\
    & SpatPlusTP2 & 0.0990 & 0.1496 & 0.3293 & 0.3462 & 0.5024 & 0.5161 \\
    \midrule
    \multirow{12}[2]{*}{\textbf{Scenario 2}} & Null  & 2.3062 & 2.3069 & 1.6992 & 1.7004 & 1.2820 & 1.2838 \\
    & Spatial & 1.8413 & 1.8477 & 1.1947 & 1.2020 & 0.7198 & 0.7344 \\
    & RSR & 2.2638 & 2.2647 & 1.6364 & 1.6382 & 1.2218 & 1.2246 \\
    & TGRMF1 & 2.1223 &	2.1278 & 1.4376 & 1.4476 & 0.9825 &	0.9956 \\
    & SpatPlus5 & 0.7342 & 0.7448 & 0.2908 & 0.3078 & 0.0231 & 0.1087 \\
    & SpatPlus10 & 0.2896 & 0.3214 & 0.2146 & 0.2480 & 0.0005 & 0.1208 \\
    & SpatPlus15 & 0.1286 & 0.1736 & 0.0208 & 0.1162 & 0.2269 & 0.2527 \\
    & SpatPlus20 & 0.0742 & 0.1387 & 0.0882 & 0.1384 & 0.2524 & 0.2736 \\
    & SpatPlusP1 & 0.4297 & 0.4476 & 0.5353 & 0.5491 & 0.6549 & 0.6665 \\
    & SpatPlusTP1 & 0.0944 & 0.1442 & 0.2271 & 0.2542 & 0.3801 & 0.3967 \\
    & SpatPlusP2 & 0.6247 & 0.6328 & 0.6542 & 0.6631 & 0.7654 & 0.7736 \\
    & SpatPlusTP2 & 0.2886 & 0.3325 & 0.4186 & 0.4481 & 0.6123 & 0.6342 \\
    \midrule
    \multirow{12}[2]{*}{\textbf{Scenario 3}} & Null  & 1.8531 & 1.8541 & 1.2042 & 1.2055 & 0.7380 & 0.7411 \\
    & Spatial & 1.4332 & 1.4384 & 0.8592 & 0.8667 & 0.3846 & 0.4034 \\
    & RSR & 1.7601 & 1.7615 & 1.1158 & 1.1178 & 0.6421 & 0.6467 \\
    & TGRMF1 &  1.2304 & 1.2352	& 0.6984 & 0.7059 &	0.2248 & 0.2493 \\
    & SpatPlus5 & 0.6002 & 0.6066 & 0.2129 & 0.2291 & 0.0630 & 0.1049 \\
    & SpatPlus10 & 0.0098 & 0.1186 & 0.0236 & 0.1227 & 0.2035 & 0.2287 \\
    & SpatPlus15 & 0.2274 & 0.2485 & 0.3499 & 0.3620 & 0.5056 & 0.5140 \\
    & SpatPlus20 & 0.3898 & 0.3985 & 0.3924 & 0.4009 & 0.5008 & 0.5078 \\
    & SpatPlusP1 & 0.5643 & 0.5720 & 0.7107 & 0.7194 & 0.7986 & 0.8060 \\
    & SpatPlusTP1 & 0.4572 & 0.4681 & 0.4465 & 0.4566 & 0.5433 & 0.5520 \\
    & SpatPlusP2 & 0.5550 & 0.5606 & 0.4987 & 0.5090 & 0.5927 & 0.5992 \\
    & SpatPlusTP2 & 0.2370 & 0.2622 & 0.1763 & 0.2308 & 0.3560 & 0.3778 \\
    \bottomrule
    \end{tabular}}%
  \label{tab:MARB_beta1}%
\end{table}%

\begin{table}[h]
  \centering
  \caption{Length of the $95\%$ credible intervals of $\beta_1$ for Simulation Study 1, Scenarios 1, 2 and 3 and $cor(\X_1,\X_2) = 0.8$, $0.5$ and $0.2$.}
    \begin{tabular}{rcccc}
    \hline\\
    & \textbf{Model} & \multicolumn{1}{c}{\textbf{cor=0.80}} & \multicolumn{1}{c}{\textbf{cor=0.50}} & \multicolumn{1}{c}{\textbf{cor=0.20}} \\
    \midrule
    \multirow{12}[2]{*}{\textbf{Scenario 1}} & Null  & 0.0628 & 0.0659 & 0.0686 \\
          & Spatial  & 0.1776 & 0.2343 & 0.2626 \\
          & RSR   & 0.0629 & 0.0672 & 0.0699 \\
          & TGRMF1 & 0.1220 & 0.1769 & 0.1983 \\
          & SpatPlus5 & 0.1485 & 0.1606 & 0.1692 \\
          & SpatPlus10 & 0.1726 & 0.1568 & 0.1588 \\
          & SpatPlus15 & 0.1614 & 0.1519 & 0.1510 \\
          & SpatPlus20 & 0.1577 & 0.1420 & 0.1413 \\
          & SpatPlusP1 & 0.1518 & 0.1423 & 0.1404 \\
          & SpatPlusTP1 & 0.1612 & 0.1490 & 0.1480 \\
          & SpatPlusP2 & 0.0810 & 0.0837 & 0.0857 \\
          & SpatPlusTP2 & 0.1028 & 0.1052 & 0.1108 \\
    \midrule
    \multirow{12}[2]{*}{\textbf{Scenario 2}} & Null  & 0.0567 & 0.0584 & 0.0603 \\
          & Spatial  & 0.3289 & 0.3548 & 0.3738 \\
          & RSR   & 0.0598 & 0.0628 & 0.0648 \\
          & TGRMF1 & 0.3269 & 0.3704 & 0.3902 \\
          & SpatPlus5 & 0.2470 & 0.2470 & 0.2483 \\
          & SpatPlus10 & 0.2752 & 0.2527 & 0.2461 \\
          & SpatPlus15 & 0.2375 & 0.2208 & 0.2161 \\
          & SpatPlus20 & 0.2243 & 0.2051 & 0.2009 \\
          & SpatPlusP1 & 0.2117 & 0.1990 & 0.1955 \\
          & SpatPlusTP1 & 0.2352 & 0.2176 & 0.2131 \\
          & SpatPlusP2 & 0.0920 & 0.0901 & 0.0919 \\
          & SpatPlusTP2 & 0.1542 & 0.1470 & 0.1492 \\
    \midrule
    \multirow{12}[2]{*}{\textbf{Scenario 3}} & Null  & 0.0487 & 0.0513 & 0.0531 \\
          & Spatial  & 0.3496 & 0.3646 & 0.3799 \\
          & RSR   & 0.0494 & 0.0524 & 0.0546 \\
          & TGRMF1 & 0.3105 & 0.3153 & 0.3265 \\
          & SpatPlus5 & 0.2434 & 0.2415 & 0.2444 \\
          & SpatPlus10 & 0.2815 & 0.2532 & 0.2479 \\
          & SpatPlus15 & 0.2501 & 0.2327 & 0.2264 \\
          & SpatPlus20 & 0.2228 & 0.2090 & 0.2070 \\
          & SpatPlusP1 & 0.2068 & 0.1973 & 0.1944 \\
          & SpatPlusTP1 & 0.2396 & 0.2282 & 0.2225 \\
          & SpatPlusP2 & 0.0665 & 0.0704 & 0.0727 \\
          & SpatPlusTP2 & 0.1025 & 0.1106 & 0.1160 \\

    \bottomrule
    \end{tabular}%
  \label{tab:length_CI}%
\end{table}%

\begin{table}[h]
        \centering
        \caption{WAIC based on 100 simulated data sets for Simulation Study 1, Scenarios 1,
2 and 3 and $cor(\X_1,\X_2) = 0.8$, $0.5$ and $0.2$.}
        \begin{tabular}{lccccccc}
        \hline\\
                & & \multicolumn{1}{c}{\textbf{cor=0.80}} & \multicolumn{1}{c}{\textbf{cor=0.50}}
& \multicolumn{1}{c}{\textbf{cor=0.20}} \\
                \midrule
                & \textbf{Model} & \multicolumn{1}{c}{\textbf{WAIC}} &
\multicolumn{1}{c}{\textbf{WAIC}} & \multicolumn{1}{c}{\textbf{WAIC}} \\
                \midrule

                \multirow{12}[2]{*}{\textbf{Scenario 1}} & Null  & 511.1920 & 570.7955 & 603.5141 \\
        & Spatial & 478.3038 & 481.9702 & 483.7083 \\
        & RSR   & 478.2911 & 481.9354 & 483.6587 \\
        & TGRMF1 & 469.1397 & 477.0736 & 479.0910 \\
        & SpatPlus5 & 467.5615 & 477.8921 & 482.1455 \\
        & SpatPlus10 & 468.1339 & 476.7807 & 482.0877 \\
        & SpatPlus15 & 467.8642 & 475.6478 & 480.7339 \\
        & SpatPlus20 & 468.5505 & 476.1541 & 481.3502 \\
        & SpatPlusP1 & 469.7314 & 478.7731 & 483.4565 \\
        & SpatPlusTP1 & 467.7620 & 476.4177 & 481.5442 \\
        & SpatPlusP2 & 512.7174 & 525.2011 & 538.8787 \\
        & SpatPlusTP2 & 512.7815 & 525.8824 & 539.8822 \\
                \midrule
                \multirow{12}[2]{*}{\textbf{Scenario 2}} & Null  & 909.9282 & 1132.7212 &
1169.1424 \\
        & Spatial & 479.9236 & 476.5743 & 477.5975 \\
        & RSR   & 479.8296 & 476.4710 & 477.4913 \\
        & TGRMF1 & 477.6394 & 477.4624 & 478.2152 \\
        & SpatPlus5 & 476.1727 & 474.9490 & 476.6245 \\
        & SpatPlus10 & 475.3869 & 474.8285 & 476.6761 \\
        & SpatPlus15 & 474.7964 & 474.5057 & 476.4105 \\
        & SpatPlus20 & 474.4604 & 473.8010 & 475.8573 \\
        & SpatPlusP1 & 474.7412 & 475.4856 & 477.2628 \\
        & SpatPlusTP1 & 474.6742 & 474.9620 & 476.7513 \\
        & SpatPlusP2 & 547.6223 & 590.7540 & 597.2066 \\
        & SpatPlusTP2 & 551.8037 & 592.1867 & 597.5115 \\
                \midrule
                \multirow{12}[2]{*}{\textbf{Scenario 3}} & Null  & 1994.5542 & 1882.3481 &
1881.4471 \\
        & Spatial & 488.7525 & 491.5679 & 493.3131 \\
        & RSR   & 488.6512 & 491.4571 & 493.1944 \\
        & TGRMF1 & 489.1019 & 491.3708 & 493.0371 \\
        & SpatPlus5 & 487.9618 & 491.2215 & 493.2403 \\
        & SpatPlus10 & 488.1447 & 490.5906 & 492.7840 \\
        & SpatPlus15 & 487.9912 & 490.4640 & 492.6054 \\
        & SpatPlus20 & 487.7714 & 490.4205 & 492.5961 \\
        & SpatPlusP1 & 488.2181 & 491.2471 & 493.1762 \\
        & SpatPlusTP1 & 488.2595 & 490.6967 & 492.7421 \\
        & SpatPlusP2 & 525.0075 & 553.3336 & 567.9812 \\
        & SpatPlusTP2 & 528.1840 & 556.9639 & 570.4615 \\
                \bottomrule
        \end{tabular}
        \label{tab:DIC_simu}%
\end{table}

\begin{table}[h]
  \centering
  \caption{Average value of mean absolute relative bias (MARB) and mean relative root mean prediction error (MRRMSE) of the relative risks based on 100 simulated data sets for Simulation Study 1, Scenarios 1, 2 and 3 and $cor(\X_1,\X_2) = 0.8$, $0.5$ and $0.2$.}
  \resizebox{\textwidth}{!}{
  \begin{tabular}{lccccccc}
  \hline\\
  & & \multicolumn{2}{c}{\textbf{cor=0.80}} & \multicolumn{2}{c}{\textbf{cor=0.50}} & \multicolumn{2}{c}{\textbf{cor=0.20}} \\
  \midrule
  & \textbf{Model} & \multicolumn{1}{c}{\textbf{MARB}} & \multicolumn{1}{c}{\textbf{MRRMSE}} & \multicolumn{1}{c}{\textbf{MARB}} & \multicolumn{1}{c}{\textbf{MRRMSE}} & \multicolumn{1}{c}{\textbf{MARB}} & \multicolumn{1}{c}{\textbf{MRRMSE}} \\
  \midrule
    \multirow{12}[2]{*}{\textbf{Scenario 1}} & Null  & 0.1506 & 0.1574 & 0.2280 & 0.2321 & 0.2622 & 0.2661 \\
    & Spatial & 0.0934 & 0.1373 & 0.0918 & 0.1561 & 0.0905 & 0.1636 \\
    & RSR & 0.0934 & 0.1373 & 0.0918 & 0.1561 & 0.0905 & 0.1636 \\
    & TGRMF1 & 0.0843 &	0.1309 & 0.0899 & 0.1550 & 0.0883 & 0.1636  \\
    & SpatPlus5 & 0.0708 & 0.1397 & 0.0833 & 0.1559 & 0.0873 & 0.1636 \\
    & SpatPlus10 & 0.0676 & 0.1525 & 0.0793 & 0.1538 & 0.0861 & 0.1616 \\
    & SpatPlus15 & 0.0606 & 0.1527 & 0.0752 & 0.1567 & 0.0836 & 0.1634 \\
    & SpatPlus20 & 0.0586 & 0.1540 & 0.0780 & 0.1577 & 0.0872 & 0.1644 \\
    & SpatPlusP1 & 0.0571 & 0.1558 & 0.0774 & 0.1637 & 0.0880 & 0.1687 \\
    & SpatPlusTP1 & 0.0591 & 0.1539 & 0.0776 & 0.1592 & 0.0868 & 0.1651 \\
    & SpatPlusP2 & 0.1078 & 0.1672 & 0.1268 & 0.1836 & 0.1404 & 0.1963 \\
    & SpatPlusTP2 & 0.1150 & 0.1666 & 0.1307 & 0.1830 & 0.1417 & 0.1952 \\
    \midrule
    \multirow{12}[2]{*}{\textbf{Scenario 2}} & Null  & 0.3893 & 0.3928 & 0.4491 & 0.4512 & 0.4736 & 0.4759 \\
    & Spatial & 0.0795 & 0.1744 & 0.0634 & 0.1682 & 0.0645 & 0.1705 \\
    & RSR & 0.0795 & 0.1744 & 0.0634 & 0.1682 & 0.0645 & 0.1705 \\
    & TGRMF1 & 0.0731 &	0.1782 & 0.0571 & 0.1744 & 0.0563 &	0.1762 \\
    & SpatPlus5 & 0.0696 & 0.1751 & 0.0591 & 0.1697 & 0.0620 & 0.1713 \\
    & SpatPlus10 & 0.0649 & 0.1791 & 0.0601 & 0.1709 & 0.0639 & 0.1716 \\
    & SpatPlus15 & 0.0644 & 0.1777 & 0.0573 & 0.1709 & 0.0610 & 0.1716 \\
    & SpatPlus20 & 0.0597 & 0.1778 & 0.0555 & 0.1702 & 0.0611 & 0.1711 \\
    & SpatPlusP1 & 0.0580 & 0.1792 & 0.0554 & 0.1754 & 0.0605 & 0.1751 \\
    & SpatPlusTP1 & 0.0606 & 0.1790 & 0.0568 & 0.1732 & 0.0614 & 0.1733 \\
    & SpatPlusP2 & 0.1100 & 0.1986 & 0.1370 & 0.2122 & 0.1454 & 0.2205 \\
    & SpatPlusTP2 & 0.1134 & 0.1993 & 0.1376 & 0.2120 & 0.1457 & 0.2203 \\
    \midrule
    \multirow{12}[2]{*}{\textbf{Scenario 3}} & Null  & 0.6622 & 0.6653 & 0.6858 & 0.6881 & 0.7195 & 0.7217 \\
    & Spatial & 0.0514 & 0.1581 & 0.0549 & 0.1610 & 0.0562 & 0.1620 \\
    & RSR & 0.0514 & 0.1581 & 0.0549 & 0.1610 & 0.0562 & 0.1620 \\
    & TGRMF1 & 0.0497 &	0.1640 & 0.0529 & 0.1667 & 0.0536 &	0.1680 \\
    & SpatPlus5 & 0.0491 & 0.1579 & 0.0537 & 0.1607 & 0.0558 & 0.1616 \\
    & SpatPlus10 & 0.0409 & 0.1619 & 0.0488 & 0.1616 & 0.0526 & 0.1614 \\
    & SpatPlus15 & 0.0399 & 0.1622 & 0.0479 & 0.1621 & 0.0525 & 0.1621 \\
    & SpatPlus20 & 0.0407 & 0.1623 & 0.0478 & 0.1619 & 0.0532 & 0.1624 \\
    & SpatPlusP1 & 0.0387 & 0.1624 & 0.0472 & 0.1631 & 0.0525 & 0.1631 \\
    & SpatPlusTP1 & 0.0410 & 0.1632 & 0.0482 & 0.1627 & 0.0531 & 0.1627 \\
    & SpatPlusP2 & 0.0904 & 0.1611 & 0.1025 & 0.1703 & 0.1141 & 0.1791 \\
    & SpatPlusTP2 & 0.0963 & 0.1613 & 0.1081 & 0.1713 & 0.1167 & 0.1795 \\
    \bottomrule
    \end{tabular}}
  \label{tab:MARB_risk}%
\end{table}%

\clearpage
\section{TGMRF}

This appendix contains the details about the marginal distributions $\F=(F_1, F_2, ..., F_{n})^{'}$ chosen for $\boldsymbol{r}=(r_1, r_2, ..., r_{n})^{'}$ and the correlation matrix $\boldsymbol{\Omega}$ that determines the spatial dependence structure in TGMRF model \citep{Prates2015}.

Any continuous distribution can be chosen as a marginal distribution of the relative risks. For instance, in this work a gamma distribution is chosen. The covariates can be incorporated either into the shape or scale parameter leading to two different gamma marginal distributions. If the covariates are included in the scale parameter,
\begin{equation*}
r_{i} \sim \Gamma(1/ \upsilon, \upsilon \exp(\X_{i, \cdot}\, \bbeta))
\end{equation*}
where $\upsilon >0$ and $\X_{i, \cdot}$ is the $i$th row of the observed covariates matrix $\X$. In contrast, if the covariates are included in the shape parameter,
\begin{equation*}
r_{i} \sim \Gamma(\exp(\X_{i, \cdot}\, \bbeta) / \upsilon, \upsilon).
\end{equation*}
Here, we consider the following priors: $\upsilon \sim \Gamma(0.01, 0.01)$ and $\beta_{i} \sim N(0, \tau=0.001)$, $j=1,\ldots, p$. As initial values for the MCMC algorithm we chose
 $\upsilon=1$ and $\beta_1=\beta_{2}= ...= \beta_{p}=0$.

An equivalent expression of the TGMRF method introduced in (\ref{TGMRF_model}) is
\begin{equation}\label{eq:TGMRF_model_2}
\boldsymbol{r} \sim TGMRF(\F, \boldsymbol{Q}_{*})
\end{equation}
which is obtained replacing the correlation matrix $\boldsymbol{\Omega}$ that determines the spatial dependence structure in the Gaussian copula by a precision matrix $\boldsymbol{Q}_{*}$. The precision matrix $\boldsymbol{Q}_{*}$ must lead to a valid correlation matrix so that $\boldsymbol{\Omega}=\boldsymbol{Q}^{-1}_{*}$.

The precision matrix of the Gaussian copula, $\boldsymbol{Q}_{*}$, is based on the precision matrix $\boldsymbol{D}-\rho\boldsymbol{M}$ of a proper conditional autorregresive (CAR) distribution. In this case, $\boldsymbol{D}$ is a diagonal matrix where the diagonal entries $d_{ii}$ are equal to the number of neighbours of the $i$th area and $\boldsymbol{M}$ is a neighbourhood matrix with non-diagonal elements $m_{ij}=1$ if areas $i$ and $j$ are neighbours and 0 otherwise. Note that $(\boldsymbol{D}-\rho\boldsymbol{M)}^{-1}$ is not a correlation matrix as its diagonal elements are not equal to one. Therefore, since copulas are scale invariant, $\boldsymbol{Q}_{*}$ is scaled as follows
%
\begin{equation*}
 \boldsymbol{Q}_{*}=\boldsymbol{\Lambda}^{1/2}(\boldsymbol{D}-\rho\boldsymbol{M})\boldsymbol{\Lambda}^{1/2},
\end{equation*}
where $\boldsymbol{\Lambda}=diag(\lambda^2_1, \lambda^2_2, ..., \lambda^2_{n})$ and $\lambda^2_{i}$ is the $i$th diagonal element of $(\boldsymbol{D}-\rho\boldsymbol{M})^{-1}$ for $i=1,2,..., n$. Then $\boldsymbol{\Omega}=\boldsymbol{Q}^{-1}_{*}$ is a correlation matrix and (\ref{eq:TGMRF_model_2}) is an equivalent way of expressing (\ref{TGMRF_model}). Here $\rho$ follows a standard uniform distribution.

\end{document}